# Unraveling Structure-Performance Trade-offs in Porous Transport Layers for PEM Water Electrolysis

Navneet Goswami[1], Sergio Diaz Abad[2], Jacob S. Spendelow[2],

Siddharth Komini Babu[2], Wilton J. M. Kort-Kamp[3,*]

[1]Energy and Natural Resources Security (EES-16), Los Alamos National Laboratory,

Los Alamos, NM 87545, United States

[2]Materials Physics and Applications Division, Los Alamos National Laboratory, Los Alamos,

NM 87545, United States

[3]Theoretical Division, Los Alamos National Laboratory, Los Alamos, NM 87545, United States

*Correspondence: kortkamp@lanl.gov

**Abstract:**

Scalable hydrogen production using proton exchange membrane water electrolyzers depends on overcoming efficiency losses arising from coupled multiphase, multicomponent transport and interfacial phenomena across the membrane electrode assembly. Here, we demonstrate a multiscale computational framework that combines pore network modeling with finite-element-based reactive transport simulations to accurately and efficiently resolve structure-performance trade-offs in porous transport layers (PTLs). We perform experiments for both commercial single-layer PTLs and microporous layer (MPL)-integrated configurations to benchmark the electrochemical model, achieving excellent agreement between modeling and measurements. We show that in single-layer PTLs, open porous networks facilitate mass transport but incur large voltage penalties from PTL-anode catalyst layer (ACL) contact resistance. Bilayer architectures with dense MPLs reduce these losses by simultaneously improving transport, contact, and structural stability. Finally, in stratified multilayer stacks, combining fine pores near the ACL with highly porous backing layers delivers superior performance at high current densities. Altogether, these results establish mechanistic guidelines for porosity-informed PTL design that minimize interfacial resistance and enable high-efficiency PEMWE operation.

## 1. Introduction

The worldwide pursuit to meet ever-rising global energy demands has necessitated the development of long-duration energy storage and flexible power-to-X technologies capable of enabling grid stabilization and sustaining large-scale industrial operations[1]. In this context, hydrogen has unlocked opportunities as a versatile energy carrier by serving as a critical feedstock for sectors such as metal refining, ammonia synthesis, and chemical production[2]. Among the various candidate hydrogen production technologies, proton exchange membrane water electrolyzers (PEMWEs) are frontrunners owing to their rapid response to fluctuating electricity, high voltage efficiency, and ability to produce high-purity hydrogen[3–6]. However, a key barrier to the widespread deployment of PEMWEs is the requirement of substantial Iridium (Ir) loading in the anode catalyst layer (ACL)[7,8], further compounded by the metal's limited abundance and its slow mining rate. Additionally, minimizing the Ir content is accompanied by a surge in losses due to poor in-plane electrical conductivity, significantly reducing catalyst utilization[9,10].

To address these issues, substantial research efforts have focused on optimizing the complex porous transport layer (PTL), which accounts for 17-25% of the component cost of a PEMWE stack[11,12]. PTLs are responsible for offering transport pathways for reactants and products and for providing suitable mechanical support for the membrane electrode assembly (MEA)[13]. The PTL material is typically composed of Titanium (Ti) to withstand harsh operating environments, including high anodic voltages and severe acidic corrosion (pH < 2)[12]. Traditionally, Ti felts and meshes have been deployed as PTL architectures, with sintered powder structures increasingly becoming popular due to recent advances in manufacturing capabilities, including powder metallurgy and additive manufacturing[14]. Over the years, researchers have formulated PTL design guidelines by targeting crucial microstructural parameters, including porosity[15,16], thickness[16–

[18]], wettability[19–21], and graded porosity[22,23]. Peng et al.[15] demonstrated that PTL bulk properties can significantly alter the mass transport resistance by regulating oxygen and water distribution pathways. Specifically, inefficient oxygen bubble removal can limit the availability of liquid water to the catalyst layer, leading to membrane dehydration and subsequently escalating mass transport and ohmic overpotentials. The optimal thickness of the PTL remains a topic of debate in the literature. Thin PTLs have been reported to enhance cell performance by reducing both ohmic resistance and gas coverage at catalytic sites[16].

A considerable performance drop arises during PEMWE operation due to the development of a compact surface oxide on Ti, which amplifies contact resistance between the PTL and the catalyst layer (CL)[24]. These interfacial challenges are further pronounced when PTLs with high porosity and pore size are employed, leading to sparse contact points with the CL and to the isolation of the catalyst particles, even in high-Ir-loaded anodes[16]. Therefore, in addition to bulk morphological properties, close attention must be paid to the PTL/CL interface topology with improved contact area, which has been shown to directly ameliorate catalyst utilization[25,26]. To mitigate the aforementioned interfacial limitations, key strategies have been explored such as (i) surface modification of the PTL through a thin coating of platinum-group metals (PGM)[27–29] such as Au, Pt, or Ir to protect against excessive passivation, and (ii) integration of a microporous layer (MPL) with the PTL[10,30–34], motivated by its successful implementation in proton exchange membrane fuel cell applications. Kulkarni et al.[31] utilized X-ray microtomography to show that incorporating an MPL increased the interfacial contact area by 20% compared to a single-layer PTL, while also augmenting oxygen removal efficiency. Schuler et al.[33] demonstrated that hierarchically structured ultra-thin Ti MPLs (≈ 20 µm) with tailored interfacial properties enhanced PEMWE performance by improving catalyst utilization, mitigating hydrogen crossover, and lowering material costs.

Complementing these experimental studies, concerted efforts have been made from a modeling perspective to probe how PTL structural characteristics govern transport processes and overall PEMWE performance[35–37]. Nouri-Khorasani et al.[38] numerically investigated oxygen bubble nucleation and growth within PTLs and evaluated their impact on the voltage penalties. They further examined how variations in CL wettability, PTL wettability, and pore size affect bubble growth, detachment, and stability behavior. More recently, pore-network models (PNM) have been extensively implemented to bridge realistic microstructural features with transport behavior, owing to their computational efficiency and ability to capture complex structure-property correlations in porous domains[39–43]. Lee et al.[39,40] applied PNM to study two-phase transport within sintered powder-based PTLs reconstructed via stochastic modeling, analyzing the effects of porosity, pore and throat sizes, and powder diameter on gas saturation and permeability. In addition to PNM, the Lattice Boltzmann Method (LBM) has been widely used to simulate multiphase flow in PTLs due to its superior capability in resolving interfacial dynamics. Peng et al.[15] systematically combined LBM with electrochemical experiments and tomography to examine the influence of morphology on oxygen distribution at different spatial locations within PTLs possessing varying porosity and mean pore size. Similarly, Paliwal et al.[44] and Satjaritanun et al.[45] employed LBM to visualize oxygen invasion patterns in the void space of PTL microstructures, further underscoring its utility for investigating microscale flow phenomena in porous media.

Based on the above discussion, it is evident that the development of next-generation composite PTLs for efficient and safe PEM water electrolysis relies on synergistic experiment-theory workflows[46] that shed light on the complex structure-transport-property relationships. This need is particularly accentuated by the challenges encountered in experimental studies, which are often limited by high costs and the complexity of processing Ti. Consequently, a systematic investigation of structure-informed PTL design via modeling is highly relevant, particularly in the regime of low anode catalyst loadings. Through a mesoscale lens, our study provides a mechanistic paradigm for how the bulk and surface characteristics of a PTL microstructure jointly influence the resulting transport-kinetics landscape, an aspect that remains insufficiently explored in the literature.

Here, we present a multiscale modeling framework that enables advanced PTL design and optimization. Firstly, we use a stochastic model to generate synthetic microstructures with prescribed morphological features (e.g., porosity, size, and orientation) representative of realistic fibrous PTLs. Next, computationally efficient PNM simulations are performed to quantify transport properties and provide porous media descriptors, such as tortuosity, effective conductivity, and permeability. We then develop a finite-element-based reactive transport model to evaluate the electrochemical performance of both single-layer PTLs and MPL-integrated PTLs with tuned pore sizes. We validate the electrochemical model against in-house experiments for both single-layer and bilayer configurations, showing good agreement. Finally, stratified architectures are considered to reveal the influence of graded porosity on transport pathways and overall device performance. To the best of our knowledge, this work combines pore-network characterization with continuum-scale electrochemical modeling for the first time to elucidate how PTL microstructural design dictates PEM electrolyzer response.

## 2. Methodology

In a typical PEMWE cell, as shown in Figure 1(a), the oxygen evolution reaction (OER) and hydrogen evolution reaction (HER) occur at the anode and cathode, respectively, liberating oxygen and hydrogen gas bubbles. To boost the efficacy of these reactions, Ir and carbon-supported platinum (Pt/C) nanoparticles are used as catalysts. The protons shuttle through the proton-exchange membrane (PEM), typically composed of perfluorosulfonic acid (PFSA); the same ionomer is also used in the anodic and cathodic catalyst layers. The MEA is compressed between a PTL on the anode and a gas diffusion layer (GDL) on the cathode.

In PEM electrolysis, the overall water-splitting reaction is expressed as

$$H_2O \rightarrow H_2 + \tfrac{1}{2}O_2, \qquad (1)$$

with the half-cell reaction at the anode side (OER) given by

$$H_2O \rightarrow \tfrac{1}{2}O_2 + 2H^+ + 2e^-, \qquad (2)$$

while the half-cell reaction at the cathode side (HER) is

$$2H^+ + 2e^- \rightarrow H_2. \qquad (3)$$

Figure 1(b) illustrates the multiscale modeling pipeline employed to generate, characterize, and correlate the complex structural features of a PTL with the electrochemical performance of a PEMWE. We initiate the workflow by creating a library of synthetically generated candidate microstructures that emulate experimentally fabricated electrospun Ti felt PTLs. This step is achieved using GeoDict[47], which enables precise control over key morphological attributes, including porosity, anisotropy, size, and shape. Next, we execute a pre-processing routine to binarize the three-dimensional structures into distinct pore and solid phases. Watershed segmentation is performed on the binarized volumes using the open-source Python package PoreSpy[48], yielding regions comprising discrete pores (nodes) and throats (local constrictions).

Information pertaining to the geometrical coordinates and connectivity of the microstructure is embedded within the extracted pore networks, forming the basis for detailed porous media transport modeling. Subsequently, simulations are accomplished on the collected networks, leveraging the capabilities of OpenPNM[49] to compute various transport properties. Mesoscale descriptors such as tortuosity, effective electronic conductivity, and single-phase absolute permeability essentially unify the underlying structure-transport relationship. The resulting effective properties are finally fed into a macrohomogeneous reactive transport model implemented in COMSOL Multiphysics[50] to evaluate the electrochemical performance of the PEMWE. This scale-bridging framework serves as an exemplar of how functional materials design can be connected to device-level performance in a high-throughput manner with minimal human intervention. In the next section, we provide details on the computational methods and equations used at each step of the workflow.

## 2.1. Synthetic structure generation

Synthetic PTL microstructures are generated using a stochastic modeling framework, an elegant approach for mimicking realistic porous architectures. Stochastic models have been widely employed in the literature for a gamut of applications, including GDLs for fuel cells[51,52], solid-state batteries[53], PTLs, and MPLs for electrolyzers[18,34,40]. Such a model offers distinct advantages to users, enabling them to explore and fine-tune a wide range of custom designs across a broad parameter space without the need to fabricate each configuration experimentally. Here, we use the FiberGeo module available in GeoDict[47], which accepts input parameters such as fiber orientation, radius, etc., and iteratively constructs the random microstructure until the target porosity is achieved. All stochastically created representative volume elements (RVEs) investigated in this work correspond to a periodic cubic domain of $250 \times 250 \times 250\ \mu m^3$. The voxel length is set to 1 µm.

## 2.2. Pore network extraction

We use a tailored watershed segmentation algorithm implemented in PoreSpy, the Sub-Network of the Oversegmented Watershed (SNOW) method developed by Gostick[54] to convert the pore volume of each generated PTL into discrete regions represented by pores connected by narrow throats. From the binarized stack corresponding to each synthetic microstructure, the extraction procedure identifies the relevant pores and throats and establishes their connectivity. A series of statistical metrics, including pore and throat diameters, volumes, size factors, throat spacing, and coordinates, is obtained using the SNOW algorithm, and isolated pores not available for transport are removed.

## 2.3. Pore network modeling

We employ the open-source pore network modeling tool (OpenPNM)[49] to calculate the transport properties. The first quantity of interest is the void-phase tortuosity, which measures the effectiveness of gas diffusion through the PTL. Tortuosity is evaluated by solving the Laplace equation, $\nabla \cdot (D\nabla c) = 0$, on the extracted pore networks, where $c$ is the concentration. To determine the tortuosity in the $x$-direction, we impose the boundary conditions $c|_{x=0} = 1\ \text{mol}/\text{m}^3$, $c|_{x=L} = 0\ \text{mol}/\text{m}^3$, where $L$ is the thickness of the PTL, and $n \cdot \nabla c = 0$ on all other faces. The diffusive flux, $J_x$, is then computed by integrating the concentration gradient, $\partial c/\partial x$, over a $y$-$z$ plane boundary. For the purpose of this calculation, the gas diffusivity in the pore network ($D$) corresponding to the void phase is considered as 1 m²/s, while that in the fiber phase is considered as 0. The $x$-direction tortuosity, $\tau_x$, is subsequently calculated using the expression below:

$$J_x = D \frac{\varepsilon}{\tau_x} \frac{c|_{x=0} - c|_{x=L}}{L}, \tag{4}$$

where $\varepsilon$ is the void volume fraction. In a similar manner, $\tau_y$ and $\tau_z$ can be computed, and finally, the mean tortuosity is given by:

$$\tau = \frac{\tau_x + \tau_y + \tau_z}{3}. \tag{5}$$

Next, we conduct simulations to evaluate the effective electronic conductivity by solving Laplace's equation for the electric potential $\nabla \cdot (\sigma_e^{bulk} \nabla \phi_e) = 0$. Analogous to the tortuosity calculation, Dirichlet and periodic boundary conditions are applied along the direction of interest and in the lateral directions, respectively. For instance, to find the effective electronic conductivity in the $x$-direction, $\sigma_{e,x}^{eff}$, we impose the following boundary conditions: $\phi_e|_{x=0} = 1$ V, $\phi_e|_{x=L} = 0$ V, and $n \cdot \nabla \phi_e = 0$ for the other orthogonal faces. For the bulk electronic conductivity, $\sigma_e^{bulk}$, we assign a value of $2.38 \times 10^6$ S/m (the intrinsic electronic conductivity of Ti[35]) to the pore network corresponding to the fiber phase and a value of 0 to the void phase. We calculate $\sigma_{e,x}^{eff}$ using the following expression:

$$\sigma_{e,x}^{eff} = \left( \int_{y-z \, plane} -\sigma_e^{bulk} \frac{\partial \phi_e}{\partial x} \bigg|_{x=0} dydz \right) \bigg/ \left( \frac{\phi_e|_{x=0} - \phi_e|_{x=L}}{L} \right). \tag{6}$$

In this study, we report the normalized electronic conductivity, $\sigma_e^*$, defined as the ratio of the mean effective electronic conductivity to its bulk value, as shown below:

$$\sigma_e^* = \frac{\sigma_{e,x}^{eff} + \sigma_{e,y}^{eff} + \sigma_{e,z}^{eff}}{3 \, \sigma_e^{bulk}}. \tag{7}$$

The third metric of interest is the single-phase permeability, which measures the efficacy of fluid flow through the PTL. The net flow rate in the pore network is first calculated by solving the mass conservation equation at each pore $(i,j)$:

$$q_i = \sum_{j=1}^{n} g_{h,ij}(p_j - p_i) = 0. \tag{8}$$

In the above equation, $i$ and $j$ indicate neighboring pores, $q_i$ represents the net flow into the pore $i$ (m³s⁻¹), and $p$ refers to the pressure at pores $i$ and $j$ (Pa). The term, $g_{h,ij}$, refers to the hydraulic conductance (m⁴skg⁻¹), determined from the Hagen-Poiseuille model for flow through a cylindrical duct, as written below:

$$g_{h,ij} = \frac{\pi d^4}{128 L_t \mu}, \tag{9}$$

where $d$ is the pore diameter (m), $\mu$ represents the dynamic viscosity of the fluid (kgm⁻¹s⁻¹), and $L_t$ is the length of the cylindrical throat (m). The overall hydraulic conductance of the pore-throat-pore assembly is calculated based on linear resistor theory using an electrical circuit analogy:

$$\frac{1}{g_{h,ij}} = \frac{1}{g_{h,pore\ i}} + \frac{1}{g_{h,throat}} + \frac{1}{g_{h,pore\ j}}. \tag{10}$$

Next, to calculate the permeability in the $x$-direction, we apply constant pressure boundary conditions at the inlet and outlet faces, such that $p_{in} = p|_{x=0} = 1$ Pa, $p_{out} = p|_{x=L} = 0$ Pa. Finally, we use Darcy's law to compute the permeability:

$$k_x = \frac{Q}{A} \frac{\mu L}{(p_{in} - p_{out})}. \tag{11}$$

In the above expression, $k_x$ represents the single-phase permeability in the $x$-direction (m²), $Q$ (m³s⁻¹) is the inlet flow rate, $\mu$ is set to unity, $A$ is the cross-sectional area (m²), and $L$ is the length of the network in the direction of flow (m). We solve Equation 8 as a set of linear equations to obtain the net volumetric flow rate through the network, $Q$. Note that the absolute (or single-phase) permeability computed from Equation 11 is not a fluid property and depends solely on the geometry of the porous medium. We can use a similar procedure delineated above to determine the permeability in the $y$ and $z$-directions to find the mean absolute permeability, given by:

$$k = \frac{k_x + k_y + k_z}{3}. \tag{12}$$

### 2.4. Equations governing the reactive transport model

The reactive transport model developed in this work accounts for the transport processes occurring in a MEA. The geometry shown in Figure 1(b), panel 4, depicts the cell sandwich configuration together with a representative finite-element mesh used for the simulations. In this framework, the gas channels integrated with the bipolar plates mathematically define the model's boundaries. The model consists of 8 equations based on the conservation of charge (electronic and protonic), mass and momentum (gas and liquid), and species (oxygen, water vapor, hydrogen, and dissolved water in the ionomer). We assume negligible crossover of gas species through the PEM. Other assumptions include steady-state and isothermal operation[55–58] of the electrolyzer in addition to ideal gas approximation[59–61]. Based on a macroscopic treatment of the porous media, we introduce the governing equations and the domains in which they are solved in the following. All relevant input parameters used in the multiphysics model are listed in Tables S5-S17 of the Supporting Information.

Conservation of electronic charge (PTL, ACL, CCL, GDL):

$$\nabla \cdot \left(-\sigma_e^{eff} \nabla \phi_e\right) = S_e. \tag{13}$$

Conservation of protonic charge (ACL, PEM, CCL):

$$\nabla \cdot \left(-\sigma_p^{eff} \nabla \phi_p\right) = S_p. \tag{14}$$

In the above equations, $\phi_e$ and $\phi_p$ are the electrostatic potentials corresponding to electron and proton transport, respectively. $S_e$ and $S_p$ represent the local electrochemical reaction source/sink terms depending on the charge carrier and electrode under consideration. The effective conductivities for the solid and the electrolyte phases are denoted as $\sigma_e^{eff}$ and $\sigma_p^{eff}$, respectively.

Conservation of oxygen species (PTL, ACL):

$$u_g \cdot \nabla c_{O_2} = \nabla \cdot \left(D_{O_2}^{eff} \nabla c_{O_2}\right) + S_{O_2}. \tag{15}$$

Conservation of water vapor species (PTL, ACL, CCL, GDL):

$$u_g \cdot \nabla c_{H_2O} = \nabla \cdot \left(D_{H_2O}^{eff} \nabla c_{H_2O}\right) + S_{H_2O}. \tag{16}$$

Conservation of hydrogen species (CCL, GDL):

$$u_g \cdot \nabla c_{H_2} = \nabla \cdot \left(D_{H_2}^{eff} \nabla c_{H_2}\right) + S_{H_2}. \tag{17}$$

The solutions of Equations (15-17) provide oxygen ($O_2$), water vapor ($H_2O$), and hydrogen ($H_2$) concentration fields. Here, $u_g$ is the gas-phase superficial velocity, and $D_i^{eff}$ (i = $O_2$, $H_2O$, and $H_2$) represents the effective diffusion coefficients of the species under consideration. The superscript, *eff,* accounts for pore blockage due to liquid water saturation and the tortuosity effect in the microstructure. The source/sink terms, $S_{O_2}$, $S_{H_2}$, $S_{H_2O}$ consider generation/depletion of the relevant species and multiphase transition effects.

Conservation of dissolved water concentration in ionomer (ACL, PEM, CCL):

$$\nabla \cdot \left(D_{DW} \varepsilon_{ionomer}^{1.5} \nabla c_{DW}\right) + \nabla \cdot \left(\frac{n_{drag} V_m c_{SO_3^-}}{F} \sigma_p^{eff} \nabla \phi_p\right) + S_{DW} = 0, \tag{18}$$

where $c_{DW}$ is the ionomer-dissolved water concentration, $D_{DW}$ is the water diffusivity in the ionomer, $\varepsilon_{ionomer}$ is the ionomer volume fraction, $n_{drag}$ is the electro-osmotic drag coefficient, $F$ is Faraday's constant, and $c_{SO_3^-}$ is the $SO_3^-$ concentration in the ionomer (sulphonic acid site density). The molar volume of the ionomer, $V_m$ can be calculated as the ratio of its equivalent weight, $EW_{ionomer}$ to the density of the dry membrane, $\rho_{dry-mem}$. It is to be noted that $\varepsilon_{ionomer} =$ 1 in the membrane and < 1 in the catalyst layers (ionomer phase in the electrodes). The source term, $S_{DW}$ further depends on the magnitude of water adsorption and desorption rates.

Conservation of liquid-phase mass (PTL, ACL, CCL, GDL):

$$\nabla \cdot (\rho_l u_l) = S_l. \tag{19}$$

Conservation of gas-phase mass (PTL, ACL, CCL, GDL):

$$\nabla \cdot (\rho_g u_g) = S_g. \tag{20}$$

Conservation of liquid-phase momentum based on Darcy's law (PTL, ACL, CCL, GDL):

$$u_l = -\frac{k k_{rl}}{\mu_l} \nabla p_l. \tag{21}$$

Conservation of gas-phase momentum based on Darcy's law (PTL, ACL, CCL, GDL):

$$u_g = -\frac{k k_{rg}}{\mu_g} \nabla p_g. \tag{22}$$

In Equations (19-22), $u_l$ and $u_g$ are the liquid and gas phase superficial velocities, while $p_l$ and $p_g$ are the liquid and gas-phase pressures, respectively. In the above expressions, $\rho_l/\rho_g$, $\mu_l/\mu_g$, and $k_{rl}/k_{rg}$ are the densities, dynamic viscosities, and relative permeabilities of the liquid/gas phases. Equations (19-22) can be further reduced to a single conservation equation in terms of pressure, as shown below:

$$\nabla \cdot \left(-\frac{\rho_l k k_{rl}}{\mu_l} \nabla p_l\right) = S_l, \tag{23}$$

$$\nabla \cdot \left(-\frac{\rho_g k k_{rg}}{\mu_g} \nabla p_g\right) = S_g. \tag{24}$$

The above equations aid in the calculation of capillary pressure $p_c$ based on the difference in liquid and gas pressures.

## 2.5. Two-phase transport

The liquid water saturation, $s$, in each domain is calculated using the Van-Genuchten water retention curve model[60] relating the saturation to capillary pressure expressed as:

$$s = \sum_{i=1}^{2} f_i [s_i(s_m - s_r) + s_r], \tag{25}$$

$$s_i = \left[1 + \left(\frac{-p_c + p^{ref}}{p_{cb,i}}\right)^{m_i}\right]^{-n_i} \text{ (anode)}, \tag{26}$$

$$s_i = 1 - \left[1 + \left(\frac{p_c + p^{ref}}{p_{cb,i}}\right)^{m_i}\right]^{-n_i} \text{ (cathode)}, \tag{27}$$

where $f_i$ are the corresponding weights; $s_r$ is the residual saturation; $s_m$ is the maximum water saturation; $p_{cb,i}, m_i, n_i$ are the Van-Genuchten fitting parameters. As stated previously, we define the capillary pressure as $p_c = p_l - p_g$ in our work, following the sign convention of García-Salaberri[60]. Once the saturation profiles of the wetting phase are obtained from Equation 25, we can readily compute transport properties, such as relative permeabilities and effective diffusivities. First, the saturation is corrected for the immobile saturation, $s_{im}$, which is assumed to be 0.1:

$$s_{red} = (s - s_{im}) / (1 - s_{im}), \tag{28}$$

where $s_{red}$ is the reduced liquid saturation. The disconnected clusters of liquid water represent the immobile phase, which does not contribute to capillary-driven transport processes. We then employ cubic power-law relationships to model the relative permeabilities of the liquid and gas phases, denoted by $k_{rl}$ and $k_{rg}$, respectively, as shown below:

$$k_{rl} = s_{red}^3, k_{rg} = (1 - s_{red})^3. \tag{29}$$

The effective diffusion coefficients of the different species (Equations 15-17) also depend on the presence of liquid water, and are given by

$$D_i^{eff} = D_{b,i} \varepsilon / \tau (1 - s_{red})^{n_b}, \tag{30}$$

where $D_{b,i}$ is the binary diffusivity of the species, $\varepsilon$ is the void volume fraction corresponding to the respective domain, $\tau$ is the tortuosity, and $n_b$ is the pore-blockage exponent. It is worth noting that within the catalyst layers, the effects of molecule-to-wall collisions are also accounted for through the Knudsen diffusion coefficients[62] (see Table S17 in the Supporting Information).

## 2.6. Membrane transport properties

The proton conductivity in the membrane and the electrode ionomer $\left(\sigma_p^{eff}\right)$ depends on the local water uptake, $\lambda$, where, $\lambda = min\left(22, \frac{c_{DW}}{c_{SO_3^-}}\right)$ and $c_{SO_3^-} = 1/V_m$. For the membrane transport properties, several correlations available in the literature have been used to quantify the effective proton conductivity, $\sigma_p^{eff}$, water diffusivity in the ionomer, $D_{DW}$, electro-osmotic drag coefficient of water, $n_{drag}$, vapor-equilibrated water content, $\lambda_{v,eq}$, and equilibrium water content, $\lambda_{eq}$ as follows[63]:

$$\sigma_p^{eff} = \varepsilon_{ionomer}^{1.5} \sigma_l^{303\,K} \exp\left[1268\left(\frac{1}{303} - \frac{1}{T}\right)\right]\,[\text{S/m}], \tag{31}$$

$$\sigma_l^{303\,K} = -7.577 \times 10^{-5}\lambda^4 + 4.24 \times 10^{-3}\lambda^3 - 8.415 \times 10^{-2}\lambda^2 + 1.138\lambda - 2.012, \tag{32}$$

$$D_{DW} = 2.5 \times 10^{-10}\left[\begin{array}{l}2.594 - 0.3371\lambda + 0.02691\lambda^2 - 6.828 \times 10^{-4}\lambda^3 \\ +1.22\exp\left(-\left(\frac{\lambda-3.045}{0.5956}\right)^2\right) - 259.4\exp(-1.686\lambda - 2)\end{array}\right] \times$$

$$\exp\left[2416\left(\frac{1}{303} - \frac{1}{T}\right)\right]\,[\text{m}^2/\text{s}], \tag{33}$$

$$n_{drag} = 2.5\frac{\lambda}{22}, \tag{34}$$

$$\lambda_{v,eq} = min(22, 0.043 + 17.81RH_{local} - 39.85RH_{local}^2 + 36RH_{local}^3), \tag{35}$$

$$\lambda_{eq} = \lambda_{v,eq} + s_{red}(22 - \lambda_{v,eq}), \tag{36}$$

$$RH_{local} = \frac{c_{H_2O} RT}{p_{H_2O}^{sat}}, \tag{37}$$

where $RH_{local}$ is the local relative humidity.

The thermodynamic expression based on the Antoine equation that relates the saturation pressure of water to the operating temperature is expressed as

$$\log(p_{H_2O}^{sat}) = 23.1963 - \frac{3816.44}{T - 46.13} \text{ [Pa]}. \tag{38}$$

### 2.7. Source terms

2.7.1. *Electrochemical reaction rates* $(R_a/R_c)$: We use Butler-Volmer kinetics to model the local volumetric reaction source/sink terms corresponding to OER and HER at the anode and cathode, respectively:

$$R_{a/c} = a_{a/c} i_{0,a/c} s^{n_s} \left[ \exp\left(\frac{2\alpha_{a/c} F}{RT} \eta_{a/c}\right) - \exp\left(\frac{-2(1 - \alpha_{a/c}) F}{RT} \eta_{a/c}\right) \right]. \tag{39}$$

In the above expression, $s^{n_s}$ represents the OER coverage factor due to liquid water access at the ACL ($n_s = 0$ in the CCL), $\alpha_{a/c}$ is the charge transfer coefficient, $a_{a/c}$ is the electrochemical active area corresponding to the catalyst layers, $\eta_{a/c}$ is the surface overpotential, $i_{0,a/c}$ is the exchange current density at the electrodes. The temperature dependence of $i_{0,a/c}$ is given by the following expression,

$$i_{0,a/c} = i_{0,a/c}^{ref} \exp\left[\frac{E_{a/c}}{R} \left(\frac{1}{T^{ref}} - \frac{1}{T}\right)\right], \tag{40}$$

where $i_{0,a/c}^{ref}$ and $T^{ref} = 353$ K denote the reference exchange current density and reference temperature. Also, $E_{a/c}$ is the activation energy for the OER/ HER half-cell reactions.

The anode and cathode overpotentials are defined as

$$\eta_a = \phi_e - \phi_p - E_a^{rev}, \tag{41}$$

$$\eta_c = \phi_p - \phi_e + E_c^{rev}, \tag{42}$$

where $E_a^{rev}$ and $E_c^{rev}$ are the reversible voltages of the OER and the HER, as expressed by the Nernst equation

$$E_a^{rev} = \frac{\Delta G_a}{2F} + \frac{RT}{4F}\log\left(\frac{p_{O_2}^{in}}{p^{ref}}\right), \tag{43}$$

$$E_c^{rev} = \frac{\Delta G_c}{2F} - \frac{RT}{2F}\log\left(\frac{p_{H_2}^{in}}{p^{ref}}\right), \tag{44}$$

with $p_{O_2}^{in}$ and $p_{H_2}^{in}$ being the partial pressures of oxygen and hydrogen in the channels, and $\Delta G_{a/c} = \Delta H_{a/c} - T\Delta S_{a/c}$ referring to the Gibbs free energy in the catalyst layers with $\Delta H_{a/c}$ and $T\Delta S_{a/c}$ representing the enthalpy and entropy variations, respectively. The global reversible cell voltage can be expressed as, $E^{rev} = E_a^{rev} - E_c^{rev}$.

2.7.2. *Water evaporation/ condensation* ($S_{ec}$): The differential between the local concentration of water vapor and the corresponding saturation value drives the phase change (water evaporation/condensation) source term:

$$S_{ec} = \begin{matrix} \gamma_e(c_{H_2O} - c_{H_2O}^{sat}), & c_{H_2O} < c_{H_2O}^{sat} \text{ (evaporation)}, \\ \gamma_c(c_{H_2O} - c_{H_2O}^{sat}), & c_{H_2O} \geq c_{H_2O}^{sat} \text{ (condensation)}, \end{matrix} \tag{45}$$

where $\gamma_e$ and $\gamma_c$ are the evaporation and condensation rate coefficients, respectively, the expressions of which are detailed below:

$$\gamma_e = k_e a_{lg} s_{red}, \tag{46}$$

$$\gamma_c = k_c a_{lg}(1 - s_{red}). \tag{47}$$

where $a_{lg}$ is the specific liquid-gas interfacial area. Also, $k_e$ and $k_c$ are the Hertz-Knudsen mass transfer coefficients, given by

$$k_{e/c} = k_{e/c}^{ref} \sqrt{\frac{RT}{2\pi M_{H_2O}}}, \tag{48}$$

with $k_{e/c}^{ref}$ as the reference evaporation/condensation mass transfer coefficient.

2.7.3. *Water adsorption/ desorption* ($S_{ad}$): The differential between the ionomer equilibrated water, $\lambda_{eq}$, and the local water uptake, $\lambda$, drives the water adsorption/desorption source term:

$$h_{ad} = \frac{k_a}{2}\left(1 - \frac{|c_{DW} - c_{SO_3^-}\lambda_{eq}|}{c_{DW} - c_{SO_3^-}\lambda_{eq}}\right) + \frac{k_d}{2}\left(1 + \frac{|c_{DW} - c_{SO_3^-}\lambda_{eq}|}{c_{DW} - c_{SO_3^-}\lambda_{eq}}\right), \tag{49}$$

$$k_{a/d} = k_{a/d}^{ref} a_{agg} \exp\left(-\frac{E_{ad}}{RT}\right), \tag{50}$$

$$S_{ad} = h_{ad}\left(c_{SO_3^-}\lambda_{eq} - c_{DW}\right). \tag{51}$$

Here, $h_{ad}$ is the rate of adsorption or desorption, $k_{a/d}^{ref}$ is the reference adsorption/desorption mass transfer coefficient, $a_{agg}$ is the agglomerate area per unit volume, and $E_{ad}$ is the activation energy for adsorption/ desorption. The domains where the different source terms are invoked are listed in Table 1.

**2.8 Experiments**

2.8.1 *MEA fabrication*: We fabricated catalyst-coated membranes (CCMs) by ultrasonic spray (Sono-tek ExactaCoat) with diluted IrO$_2$ (Alfa Aesar Premion @ 99.99% purity) for the anode and Pt/C (TEC10V50E, Tanaka) for the cathode. The catalyst particles were dispersed in an ultrapure DI water (Millipore, 18.2 ohm-cm) and n-propanol (Sigma-Aldrich) mixture in a 4:3 ratio. D2020 was used as a binder, with an ionomer-to-catalyst ratio of 0.1 for the anode and 0.5 for the cathode. We sprayed catalyst inks in N115 held at 87 °C on a vacuum plate in a Sono-tek ExactaCoat ultrasonic spray system with a 25 kHz Accumist nozzle. The anode catalyst loading was fixed as 0.1 mg$_{Ir}$cm$^{-2}$, while the cathode catalyst loading was fixed in all cases as 0.1 mg$_{Pt}$cm$^{-2}$. X-ray Fluorescence (XRF) was used to control and confirm the catalyst loading after spray coating.

2.8.2 *PEMWE assembly and operation*: We used a 50 cm$^2$ hardware unit equipped with triple serpentine flow fields in this work (Fuel Cell Technologies), masked down to 5 cm$^2$ using PTFE gaskets of specific thickness to achieve 10% compression in the anode and cathode. A carbon GDL (MGL370, Avcarb) with 78% porosity and 370 $\mu$m thickness was used for the cathode. Platinum-coated Ti felt (2GDL06 and 2GDL10-0.25 BS02PT, Bekaert) with a thickness of 250 $\mu$m was employed as PTL for the anode. Prior to assembly, the MEA was immersed in ultrapure DI water for at least 4 h, and pre-heated water at 80 °C was left flowing through the anode (10 mL min$^{-1}$) for another 4 h after the cell reached 80 °C.

2.8.3 *Electrochemical characterization and degradation measurements*: Polarization curves and electrochemical impedance spectroscopy (EIS) measurements were performed before and after conditioning. The conditioning protocol consisted of a first step of square-wave cycling (500 cycles) from 0.6 V to 1.6 V, holding each potential for 2.5 s, and a second step in which the current was held at 2 A/cm$^2$ for 3 hr. Performance was measured using a 3-way polarization curve from 0 to 4 A/cm$^2$, consisting of a first anodic scan, a cathodic scan, and a second anodic scan. Only the second anodic scan was used for comparison purposes. This 3-way polarization curve eliminates variation at low current densities caused by metallic iridium phases formed during the conditioning protocol. High-frequency resistance (HFR) was measured at each point on the polarization curve using EIS with a current amplitude of 2% up to 0.2 A/cm$^2$, then fixed at 0.2 A, over a frequency range of 10 Hz to 40 kHz.

## 3. Results and Discussion

### 3.1. Unifying structure-transport relationships

The stochastically generated PTL microstructures for different porosities in the range of 0.35-0.75 are displayed in Figure 2(a). Note that the candidate structures are isotropic, with fibers randomly oriented. In addition, we have considered a monodisperse distribution of fiber size, with a fixed radius of 5 μm. After obtaining the binarized phase distribution (voids and solids) and its connectivity, we perform the pore network extraction step as described in Section 2.2. Figure 2(b) illustrates the void-phase pore network overlaid on the fiber structure (left) and the standalone network (right) corresponding to the low-porosity microstructure ($\varepsilon_{PTL} = 0.35$). The pore network visualization captures the interconnected nature of the void channels responsible for gas diffusion. Similarly, visualization of the fiber-phase pore network for the high porosity case ($\varepsilon_{PTL} = 0.75$) is presented in Figure 2(c), showing the solid skeleton that aids electron transport. The pore size distributions (PSDs) and throat size distributions (TSDs) for these extreme porosity configurations are shown in Figures 2(d–e). As porosity increases, the mean pore and throat diameters shift toward larger values; quantitatively, the mean pore size is 13 μm for the low-porosity structure, increasing to 22.5 μm for the high-porosity structure.

Figures 3(a-b) compare the tortuosities ($\tau$) and normalized effective electronic conductivities ($\sigma_e^*$) predicted by PNM with benchmark finite difference method (FDM) calculations performed using PoreSpy[48]. In reality, pores and throats can exhibit different shapes, and accordingly, the diffusive and hydraulic conductances are corrected using the concept of size factors in OpenPNM. Here, we calibrate our pore network model by assigning the size factors as 'pyramids and cuboids' for the void phase and 'cones and cylinders' for the fiber phase. The maximum error for $\tau$ and $\sigma_e^*$ (across the entire porosity range) is approximately 4.1% and 12.2%, respectively, ascertaining the reliability of the PNM model. Next, we screen a combination of PTL morphological attributes, including porosity and fiber radius, to examine their influence on key transport properties, as shown in Figures 3(c-e). As seen in Figure 3(c), for $r_{fiber} = 5$ μm, $\tau$ changes from 2.85 at $\varepsilon_{PTL} = 0.35$ to 1.24 at $\varepsilon_{PTL} = 0.75$, consistent with an open void network. Similarly, going from $\varepsilon_{PTL} = 0.35$ to 0.75, we note that $\sigma_e^*$ drops from 0.40 to 0.07 in Figure 3(d). This observation stems from resistance to electron percolation due to diminished connectivity of the solid matrix at lower fiber packing fractions. Neither of these properties reflected a distinct behavior with the variation in fiber radius. However, in stark contrast, the trend in absolute permeability, Figure 3(e), reveals a strong dependence on fiber size, in addition to PTL porosity. This is commensurate with prior works in the literature, which suggested that the single-phase permeability, $k$, scales with $r_{fiber}^2$ for fibrous microstructures[64].

### 3.2. Model-experiment synergy and analysis of single-layer PTLs

Following the workflow outlined in Figure 1(b), we incorporate the computed effective properties (see Table 2) into the reactive transport model to assess the electrochemical landscape ensuing from changes in the PTL design. The electrochemical model validation is first established by comparing the simulated polarization response with experimental polarization curves for MEAs with commercially fabricated PTLs (2GDL06 and 2GDL10) in both single-layer and bilayer configurations, as shown in Figures 4(a-b). We set the operating conditions for the experiment at a balanced pressure of 1 atm and a temperature of 80 °C. Ultra-low catalyst loadings are considered for both electrodes (~0.1 mg/cm$^2$), as explained in section 2.8.1. Note that the porosity of the 2GDL10 and 2GDL06 PTLs is 0.56 and 0.74, respectively, with a thickness of 250 μm. The corresponding values of the GDL porosity and thickness are 0.78 and 370 μm, respectively. For the single-layer case, Figure 4(a), the contact resistance at the PTL-ACL interface has been phenomenologically captured in our model using a Bruggeman-type approximation[19],

$$R_{interfacial, PTL-ACL} = K_{PTL-ACL}\varepsilon_{PTL}^{1.5}, \qquad (52)$$

where, $K_{PTL-ACL}$ is a proportionality constant referred to as the specific interfacial resistance in this work. We treat the parameter, $K_{PTL-ACL}$ as a fitting constant and assume it remains invariant across the porosity window considered. Based on the above treatment, we observe that the model accurately predicts the experimental polarization behavior across the entire current density range. The best fit is obtained with $K_{PTL-ACL} = 80$ mΩ cm².

Next, we present the model validation for the bilayer PTL architecture in Figure 4(b). The inset schematically depicts the bilayer arrangement on the anode side of the PEMWE, where the PTL is integrated with an MPL. The porosity and thickness of the MPL are 0.105 and 15 μm, respectively, with the relevant effective properties calculated as, $\tau_{MPL} = 30.553, \sigma_e^* = 0.872$, and $k = 9.54 \times 10^{-16}$ m². For this analysis, we invoke the following interfacial mechanisms to account for potential resistive effects induced by the MPL. Contact resistance at the MPL-ACL interface:

$$R_{interfacial, MPL-ACL} = K_{MPL-ACL}\varepsilon_{MPL}^{1.5}, \qquad (53)$$

and porosity mismatch at the PTL-MPL interface:

$$R_{interfacial, PTL-MPL} = K_{PTL-MPL}|\varepsilon_{PTL} - \varepsilon_{MPL}|^{1.5}. \qquad (54)$$

The reasonable agreement demonstrates the model's ability to capture the experimental trend, albeit with minor deviations observed in the 2GDL06 case coupled with the MPL in the kinetically limited regime (below 0.5 A/cm²). The best fit is obtained with $K_{MPL-ACL} = 50$ mΩcm² and $K_{PTL-MPL} = 0.2K_{MPL-ACL}$. Importantly, the marginal effect of the PTL backing layer on cell performance is consistent with observations by Jung et al.[10], who recently reported a limited role of the same in dictating the net resistance for dual-layer PTLs. Additionally, for this configuration, the anode exchange current density $(i_{0,a})$ is $2.5 \times 10^{-2}$ A/m², which is about three times higher than the value of $8 \times 10^{-3}$ A/m² for the baseline (single-layer) scenario. We believe that a smoother, more uniform contact distribution between the MPL and ACL facilitates greater catalyst utilization and, consequently, promotes reaction kinetics. Moreover, the specific interfacial resistance in contact with the ACL (50 mΩ cm²) is lower than that reported in the single-layer configuration (80 mΩ cm²), further indicating improved interfacial contact due to the presence of the MPL.

Once the model's predictive capability is ensured, we perform further analysis to delineate the influence of a key design parameter, i.e., the porosity of a single-layer PTL (at a fixed fiber radius of 5 µm) on electrolyzer performance in Figures 5(a-e). Figure 5(a) reveals that higher cell voltages occur as porosity increases from 0.35 to 0.75, particularly at high operating current densities. We decouple the associated voltage loss components at the highest current density, $I = 4$ A/cm² (shown in Figures 5(b-c)), following the procedure outlined in Section S1 of the Supporting Information. A slight drop in the kinetic loss appears when the porosity increases ($\eta_{kinetic} = 0.462$ V at $\varepsilon_{PTL} = 0.35$ versus $\eta_{kinetic} = 0.451$ V at $\varepsilon_{PTL} = 0.75$). On the other hand, the HFR comprising the ohmic loss contributions from proton and electron conduction in the catalyst layers, PTL, and GDL, rises considerably towards higher porosity levels ($\eta_{ohmic} = 0.297$ V at $\varepsilon_{PTL} = 0.35$ versus $\eta_{ohmic} = 0.444$ V at $\varepsilon_{PTL} = 0.75$). The reason behind the surge in the ohmic loss emanates from increased contact resistance at the PTL-ACL interface, consistent with Equation 52. Meanwhile, as we progressively increase the PTL porosity, the mass transport loss depicted in Figure 5(c) monotonically decreases, indicating an improved extent of gas diffusion at $\varepsilon_{PTL} = 0.75$ ($\eta_{mt} = 0.028$ V at $\varepsilon_{PTL} = 0.35$ versus $\eta_{mt} = 0.019$ V at $\varepsilon_{PTL} = 0.75$). To shed further light from a mechanistic perspective, we next highlight the cross-sectional (in-plane) averaged profiles of relevant species, illustrating their through-plane transport behavior. Our observation pertaining to the variation of mass transport loss is corroborated by the oxygen concentration profiles along the anode (Figure 5(d)), which suggests that low porosity PTLs lead to oxygen accumulation, which is pronounced near the front of the anode catalyst layer (in proximity to the ACL-membrane interface, highlighted by the vertical dashed line at the extreme left). This signature of oxygen buildup can be attributed to the smaller pore sizes and tortuous pathways, which hinder the removal of electrogenerated oxygen bubbles within the ACL via the PTL. On the other hand, the hydrogen throughput on the cathode side (shown in dashed lines on the secondary $y$-axis) remains invariant, which is an anticipated behavior. In Figure 5(e), we see that the average liquid water saturation along the PTL thickness increases with PTL porosity, reflecting reduced hydraulic resistance and improved ingress of reactant water through the open void regions. The limited water accessibility faced by PTLs with low porosity reduces the effective electrochemically active area (Equation 39), further explaining the modest increase in the kinetic loss for densely packed PTLs. Overall, the multiphysics analysis of a single-layer architecture demonstrates a key trade-off between gas removal/water supply and the dominant ohmic loss, underscoring the need for an optimized PTL design that prioritizes the PTL-ACL contact resistance responsible for electron conduction, while ensuring balanced transport between reactants and products.

## 3.3. Mechanistic interrogation of multilayer architectures

In Figure 6, we focus on developing a comprehensive mechanistic understanding of how the confluence of interfacial and bulk features in a bilayer PTL affects the polarization response of a PEMWE. It is worth noting that we have applied the same porosity window and corresponding effective porosities of the PTL (displayed in Table 2) also to the MPL, along with the bulk electronic conductivity associated with Ti. Furthermore, we have chosen a porosity of 0.75 for the base PTL (backing layer). The rationale for using such a backing layer is that commercially fabricated PTLs are typically highly porous, which compromises contact with the ACL and exacerbates ohmic overpotentials. For the subsequent simulation results shown in Figure 6(a), we have considered the thicknesses of the PTL and the MPL fixed at 250 µm and 15 µm, respectively. Now, for the validation analysis reported in Figure 4(b), we obtained an optimal fit to the experimental polarization curve pertaining to a ratio of $K_{PTL-MPL}/K_{MPL-ACL} = 0.2$. In Figure 6(a), we extend this analysis by systematically varying the specific interfacial resistance ratio (i.e., $K_{PTL-MPL}/K_{MPL-ACL}$) to explore regimes that are difficult to realize experimentally, while simultaneously portraying the crucial role of property mismatch in governing the concomitant transport phenomena. Mechanistically, a zero value of the interfacial resistance ratio ($K_{PTL-MPL}/K_{MPL-ACL} = 0$) reflects a stepped single-layer microstructure featuring a discrete porosity transition, with negligible resistive contribution from the PTL-MPL interface. On the other hand, progressively increasing values of the ratio signify a more thoroughly layered PTL. It is visible that the cell voltage at 4 A/cm² is least for dense MPLs in a region where the overall contact resistance is minimized ($K_{PTL-MPL}/K_{MPL-ACL}$ approaching 0). Notably, a difference in voltage of ~86 mV is observed when comparing the extreme cases ($K_{PTL-MPL}/K_{MPL-ACL} = 0$, $\varepsilon_{MPL} = 0.35$ versus $K_{PTL-MPL}/K_{MPL-ACL} = 1$, $\varepsilon_{MPL} = 0.75$), further reinforcing the importance of rigorously engineering interfaces for better performance. Low MPL porosities result in oxygen confinement at the anode ($c_{O_2,average} = 119.25$ mol/m³), raising the mass transport resistance (refer to Figure S2 of the Supporting Information). Conversely, an MPL with an open, well-connected pore network ($\varepsilon_{MPL} = 0.75$) ensures oxygen evacuation from the CLs, alleviating transport limitations ($c_{O_2,average} = 108.14$ mol/m³). Notably, we identify that the interfacial resistance ultimately dictates overall cell performance, dominating the mass-transport contributions by nearly an order of magnitude. To further understand the combined interplay of MPL porosity and thickness on the ensuing interfacial complexations, we present a phase map of the cell voltage at 4 A/cm$^2$ in Figure 6(b). A dashed line bounded within 1.94 V is used to demarcate the optimal operation zone, which corresponds to a region of MPL porosity $\varepsilon_{MPL} \approx 0.35 - 0.5$ and MPL thickness $L_{MPL} \approx 15 - 55$ µm. This phase map indicates that cell voltage is less sensitive to changes in MPL thickness, thereby emphasizing the need to adopt porosity-informed PTL design strategies. Moderately dense MPLs are found to be most preferable for robust PEMWE operation, as ultra-dense MPLs may impede mass transport (refer to Figure S3 of the Supporting Information), while ultra-porous architecture comes at the expense of reduced interfacial contact and structural integrity. We would like to highlight at this point that the results demonstrated in Figure 6(b) were quantified at $K_{PTL-MPL}/K_{MPL-ACL} = 1$; however, other values of this ratio would also reveal similar characteristic features.

We present the influence of PTL layer stratification on the electrochemical performance of the PEMWE in Figure 7. While introducing multiple layers poses experimental challenges as it necessitates precise control over additional interfaces, this remains a promising direction. For our analysis, we consider a tri-layer PTL, with the backing layer porosity (PTL) set to 0.75 and the MPL porosity to 0.45 (optimal zone in Figure 6(b)). First, we vary the porosity of the middle layer (referred to as the interlayer, IL) to assess its impact on cell performance, as depicted in Figure 7(a). Overall, we observe that an intermediate IL porosity, $\varepsilon_{IL} = 0.65$ delivers optimal performance, indicating the indispensable need to maintain a uniform porosity gradient throughout the multilayer stack. The breakdown of the corresponding voltage-loss modes is shown in Figure 7(b), which reveals that the kinetic contribution remains essentially unchanged across the range of interlayer porosities, while the ohmic loss exhibits a minimum at $\varepsilon_{IL} = 0.65$. On the other hand, the mass transport loss displayed in Figure 7(c) is relatively small compared to the kinetic and ohmic loss terms and decreases slightly with increasing IL porosity, corroborated with improved gas diffusion features through a more open interlayer network. Next, in Figure 7(d), we compare the electrochemical response of three PTL architectures, single-layer, bilayer, and trilayer, across selected current densities. At $I = 0.1$ A/cm$^2$, we notice that the two-layered configurations exhibit nearly identical voltages (1.428 V). The contrast across the three scenarios tends to become more pronounced as we increase the operating current density. For instance, at current densities such as 0.5 and 1 A/cm$^2$, the highest voltage is registered in the case of the single-layer PTL; while the layered configurations lower the cell voltage by several millivolts to provide enhanced performance owing to improved catalyst utilization and reduced interfacial resistance, as elaborated in Figure 4(b). At 4 A/cm$^2$, the divergence is most visible, with 2.084 V for the single-layer, 1.937 V for the bilayer, and 1.928 V for the trilayer, demonstrating that stratification of the PTL layer progressively lowers the cell voltage, with the benefit extracted most significantly at high current densities. Figures 7(e-f) further examine the role of the porosity gradient direction in controlling the performance. In the "decreasing-porosity" microstructure displayed in Figure 7(e), the porosity reduces toward the anode catalyst layer (75% → 65% → 45%), whereas in the "increasing-porosity" design, it rises in the same direction (45% → 65% → 75%). The corresponding polarization curves in Figure 7(f) highlight that the decreasing-porosity architecture consistently yields lower cell voltages across the full current density range (~71 mV at 4 A/cm$^2$). This behavior indicates that positioning finer pores adjacent to the ACL is advantageous for overall cell performance, as it enhances oxygen bubble removal efficiency[22] and electron conduction via uniform conformal contact.

## 4. Conclusions

This study presents a novel, multiscale model for the predictive design and optimization of PTLs in PEM water electrolysis. Stochastically generated microstructures mimicking electrospun fiber-based PTLs were created by varying key morphological attributes, including porosity and fiber size. Pore-network modeling was then applied to these microstructural realizations to assess structure-transport relationships in a high-throughput manner. The resulting effective properties showed a strong dependence on PTL packing fraction, with the single-phase absolute permeability additionally scaling with fiber size. Mesoscale simulations were performed using an experimentally benchmarked reactive transport model to shed light on the ensuing electrochemical landscape. Deconvolution of voltage losses in single-layer PTLs revealed the dominant effect of PTL-ACL contact resistance, emphasizing the critical need to tailor interfacial mechanisms to precisely mitigate voltage penalties. Our mechanistic interrogation of bilayer PTLs suggested integrating moderately dense MPLs to balance the trade-offs among mass transport, interfacial contact, and mechanical stability. The interplay of interfacial property mismatch and porosity in dictating performance and oxygen buildup behavior was also elucidated. Furthermore, stratified PTLs with controlled, uniform porosity gradients delivered enhanced cell performance, particularly at high operating current densities. Future directions include developing physics-based sub-models to mechanistically predict and characterize interfacial contact resistance for direct integration into the current computational workflow. Concerted efforts should also be directed toward examining the role of pore size distributions in capturing two-phase transport metrics, such as capillary pressure-saturation relationships, rather than relying solely on power-law descriptions. Overall, the present work underscores the strong coupling between morphology and bulk transport, paving the way for digital twins that establish strategic guidelines for structure-informed PEMWE design.

**Conflicts of Interest**

The authors declare no conflicting interests.

**Data availability**

The data supporting this article are included in the Supporting Information.

**Acknowledgments**

Research was supported by the Laboratory Directed Research and Development (LDRD) program of Los Alamos National Laboratory under project number 20240061DR.

**Table 1.** Source terms corresponding to the conservation equations solved in the reactive transport model

| Source term | PTL | ACL | PEM | CCL | GDL |
|---|---|---|---|---|---|
| $S_l$ | $M_{H_2O}S_{ec}$ | $M_{H_2O}S_{ec}$ | — | $M_{H_2O}S_{ec}$ | $M_{H_2O}S_{ec}$ |
| $S_g$ | $-M_{H_2O}S_{ec}$ | $-M_{H_2O}(S_{ec}+S_{ad}) + M_{O_2}\dfrac{R_a}{4F}$ | — | $-M_{H_2O}(S_{ec}+S_{ad}) + M_{H_2}\dfrac{R_c}{2F}$ | $-M_{H_2O}S_{ec}$ |
| $S_e$ | 0 | $-R_a$ | — | $R_c$ | 0 |
| $S_p$ | — | $R_a$ | 0 | $-R_c$ | — |
| $S_{DW}$ | — | $S_{ad}-\dfrac{R_a}{2F}$ | 0 | $S_{ad}$ | — |
| $S_{O_2}$ | 0 | $\dfrac{R_a}{4F}$ | — | — | — |
| $S_{H_2}$ | — | — | — | $\dfrac{R_c}{2F}$ | 0 |
| $S_{H_2O}$ | $-S_{ec}$ | $-S_{ec}-S_{ad}$ | — | $-S_{ec}-S_{ad}$ | $-S_{ec}$ |

**Table 2.** PNM-predicted effective properties used as input parameters in the reactive transport model

| $\varepsilon_{PTL}$ | **0.35** | **0.45** | **0.55** | **0.65** | **0.75** |
|---|---|---|---|---|---|
| $\tau$ | 2.848 | 2.061 | 1.641 | 1.401 | 1.242 |
| $\sigma_e^*$ | 0.395 | 0.280 | 0.191 | 0.121 | 0.068 |
| $k$ (m²) | $2.052\times 10^{-13}$ | $5.915\times 10^{-13}$ | $1.421\times 10^{-12}$ | $2.890\times 10^{-12}$ | $5.295\times 10^{-12}$ |

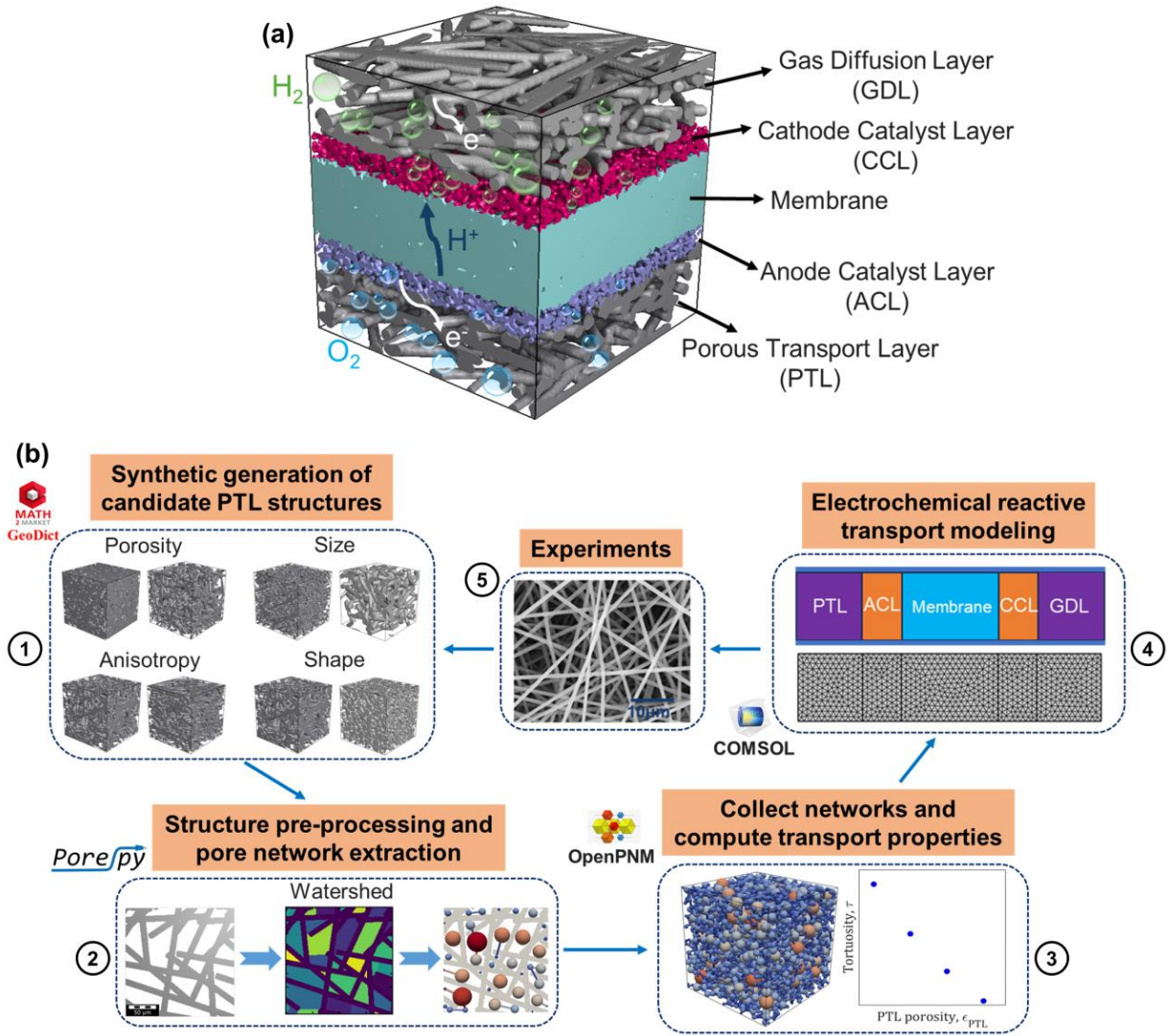

**Figure 1:** (a) Schematic of a PEM water electrolyzer. (b) Workflow of the multiscale modeling framework illustrating the synthetic PTL microstructure generation and the exchange of information between the pore-network model and the macrohomogeneous electrochemical reactive transport model.

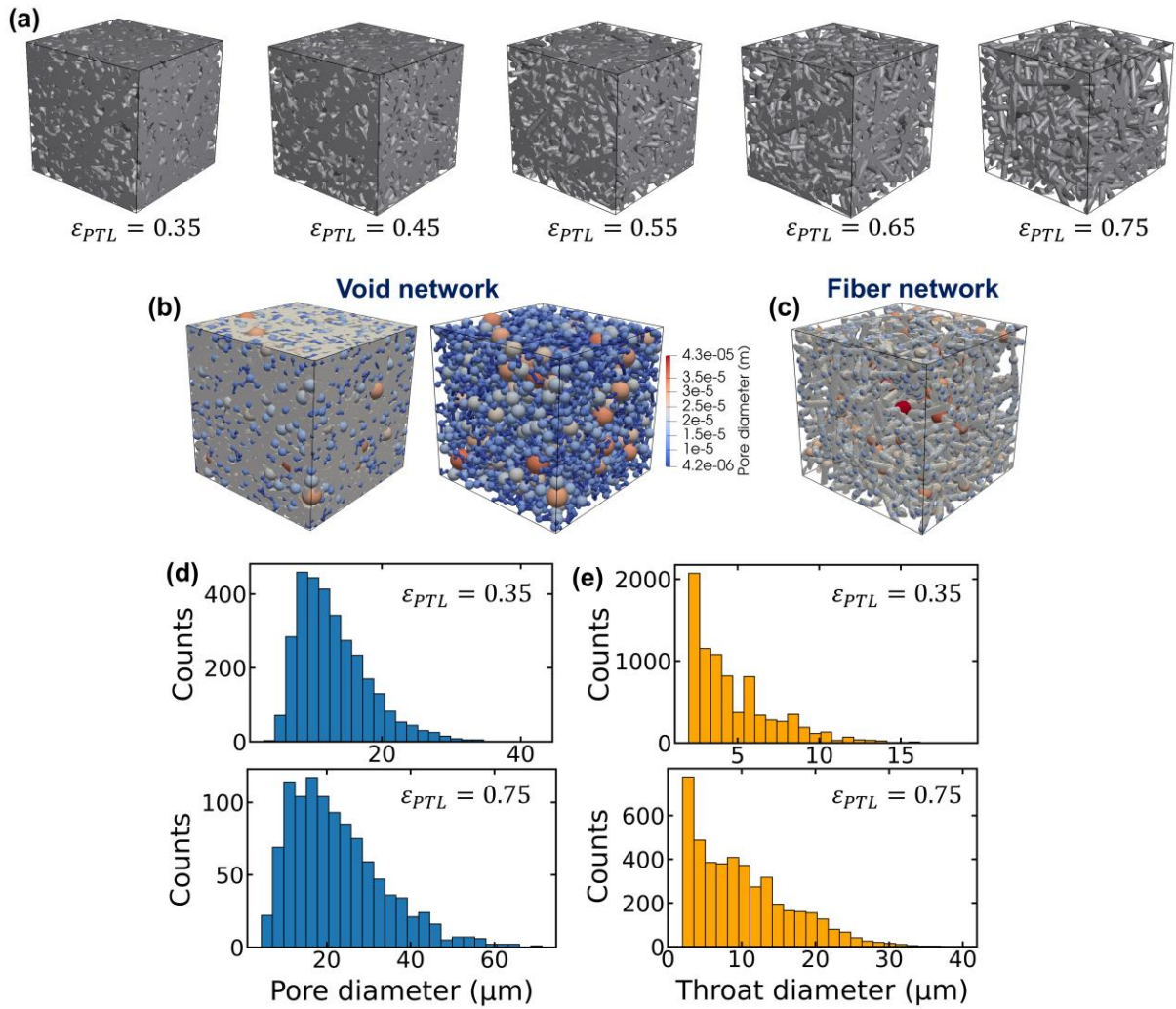

**Figure 2:** (a) Stochastically generated PTL structures with fibrous morphology at different porosity levels. (b) Void-phase pore network for $\varepsilon_{PTL} = 0.35$, shown both overlaid on the fiber structure (left) and as a standalone network (right). (c) Fiber-phase pore network for $\varepsilon_{PTL} = 0.75$, overlaid on the corresponding structure. (d) Pore size distribution (PSD) for $\varepsilon_{PTL} = 0.35$ and $0.75$. (e) Throat size distribution (TSD) for $\varepsilon_{PTL} = 0.35$ and $0.75$.

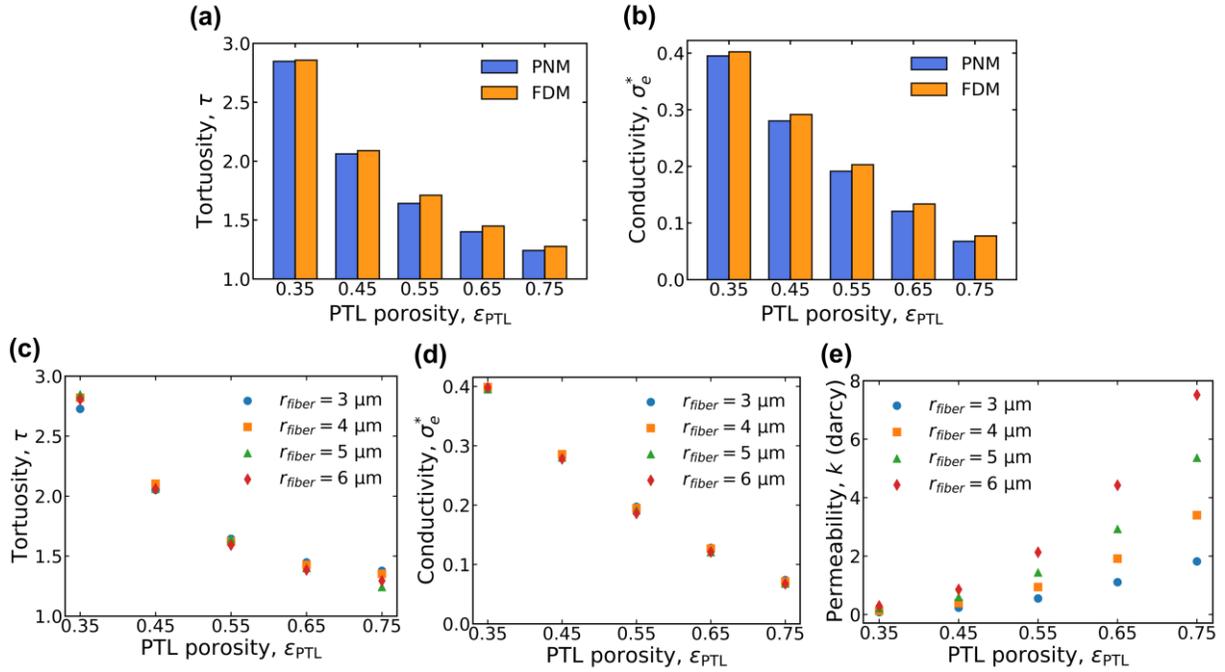

**Figure 3:** (a) Comparison of PNM-predicted tortuosity with corresponding finite difference method (FDM) calculations. (b) Comparison of PNM-predicted normalized effective electronic conductivity (Equation 7) with FDM-based results. Influence of PTL porosity ($\varepsilon_{PTL}$) and fiber radius ($r_{fiber}$) on (c) tortuosity ($\tau$), (d) normalized effective electronic conductivity ($\sigma_e^*$), and (f) single-phase (absolute) permeability ($k$).

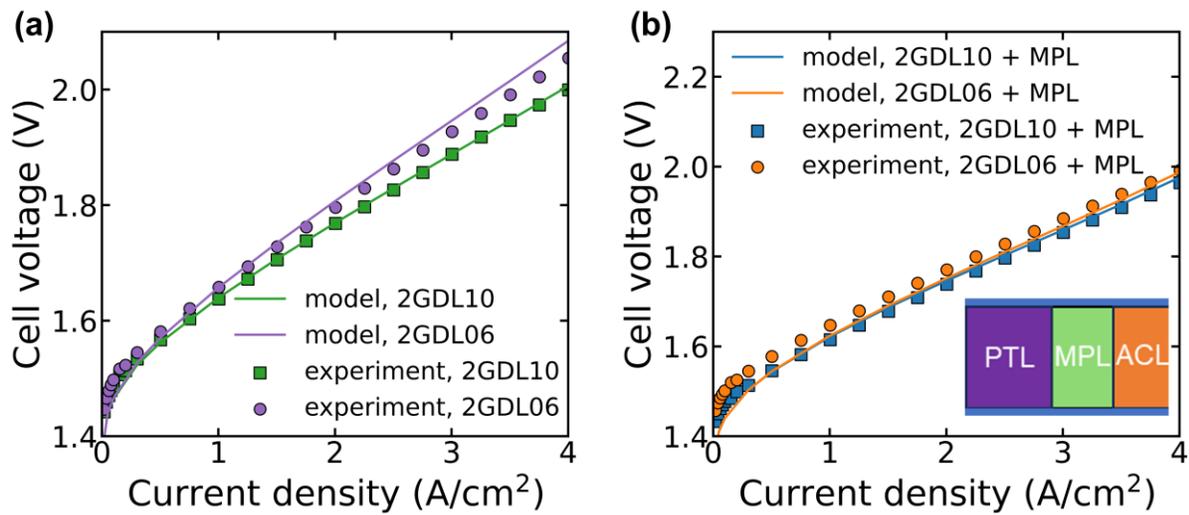

**Figure 4:** Validation of the electrochemical model against experimental polarization curves at ultra-low loadings of 0.1 mg/cm$^2$, balanced pressure of 1 atm, and an operating temperature of 80 °C for (a) single-layer configuration with commercially fabricated Bekaert 2GDL10 and 2GDL06 PTLs, (b) bilayer configuration with 2GDL10 and 2GDL06 PTLs integrated with a microporous layer (MPL) of thickness 15 μm and 10.5% porosity.

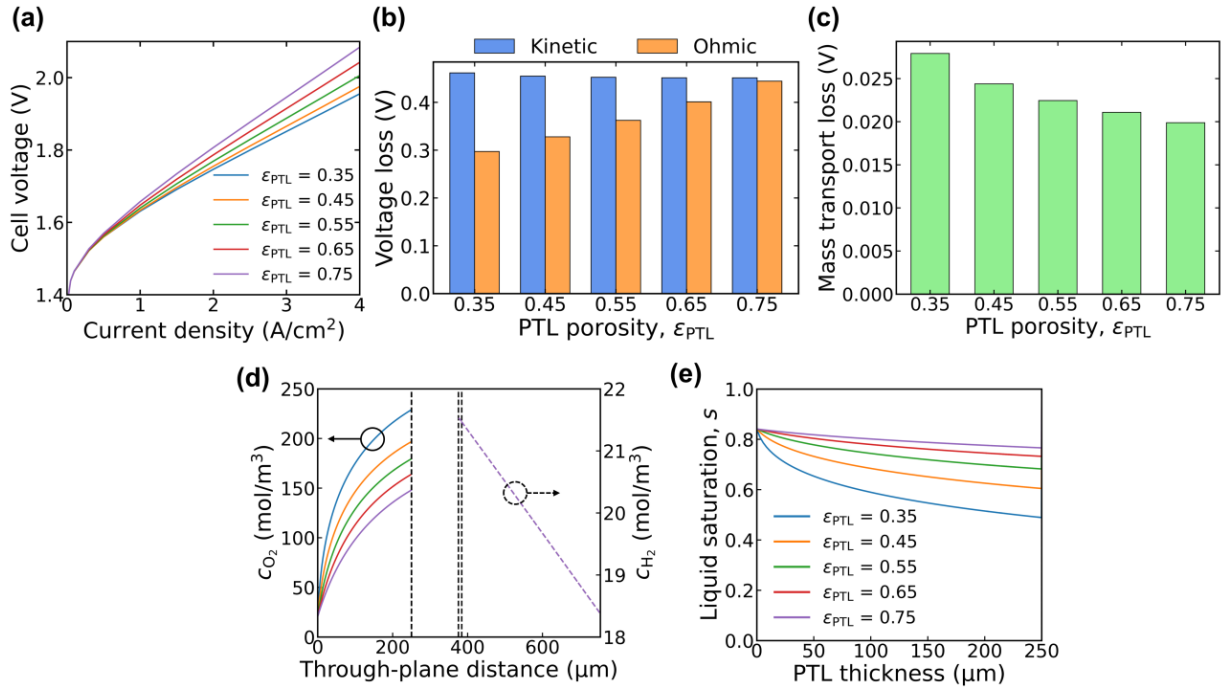

**Figure 5**: Effect of single-layer PTL on electrochemical performance: (a) polarization curves displaying the influence of porosity, along with the associated voltage-loss deconvolution at $I = 4$ A/cm$^2$ - (b) kinetic and ohmic losses, and (c) mass transport loss. (d) Oxygen and hydrogen concentration profiles along the through-plane direction of the membrane electrode assembly at $I = 4$ A/cm$^2$. (e) Liquid water saturation profiles along the PTL thickness at $I = 4$ A/cm$^2$.

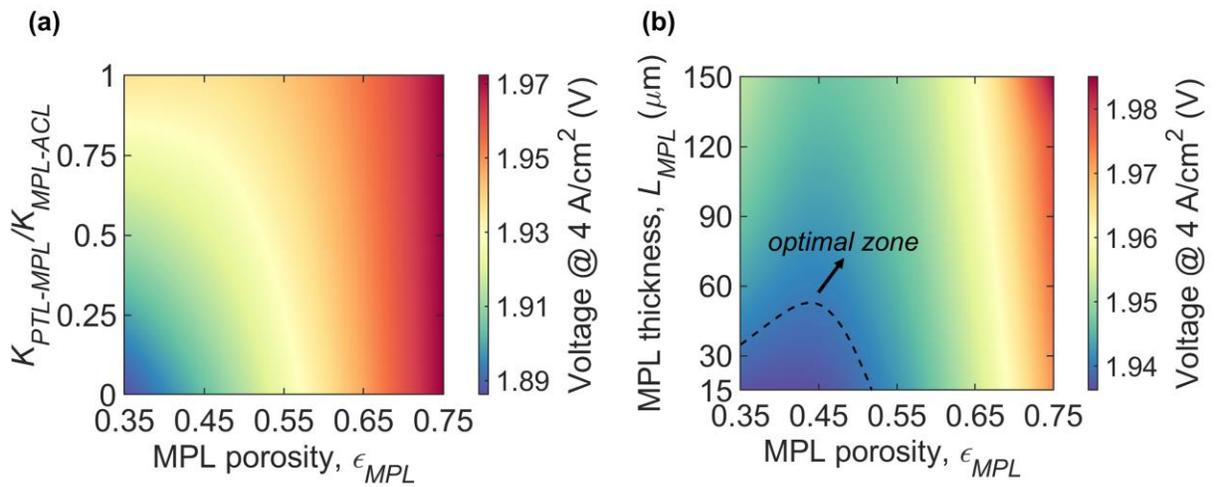

**Figure 6**: Effect of bilayer PTL on electrochemical performance: (a) interplay of the specific interfacial resistance ratio and MPL porosity on the cell voltage at $I$ = 4 A/cm$^2$, and (b) interplay of MPL porosity and thickness on the cell voltage at $I$ = 4 A/cm$^2$, with the PTL backing-layer porosity fixed at 0.75.

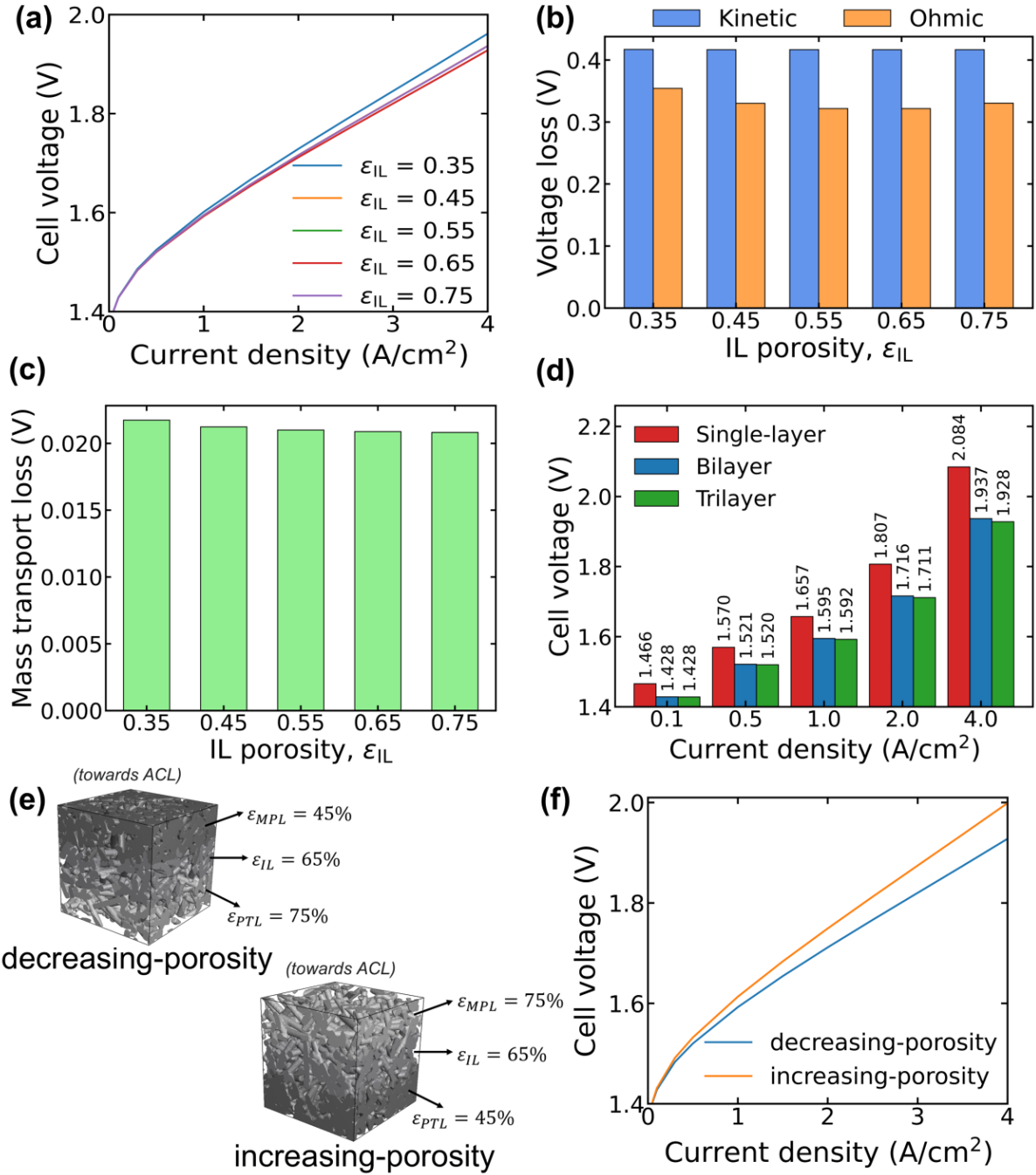

**Figure 7:** Effect of PTL layer stratification on electrochemical performance: (a) polarization curves displaying the influence of interlayer (IL) porosity for trilayer PTLs, along with the associated voltage-loss deconvolution - (b) kinetic and ohmic losses, and (c) mass-transport loss. (d) Performance comparison of single-layer, bilayer, and trilayer PTL architectures at selected current densities. (e) Microstructures illustrating decreasing and increasing porosity gradients within stratified PTLs, with (f) corresponding performance comparison showing the influence of porosity-gradient direction on overall cell performance.

# Supporting Information

**Unraveling Structure-Performance Trade-offs in Porous Transport Layers for PEM Water Electrolysis**

Navneet Goswami[1], Sergio Diaz Abad[2], Jacob S. Spendelow[2],

Siddharth Komini Babu[2], Wilton J. M. Kort-Kamp[3,*]

[1]Energy and Natural Resources Security (EES-16), Los Alamos National Laboratory,

Los Alamos, NM 87545, United States

[2]Materials Physics and Applications Division, Los Alamos National Laboratory, Los Alamos,

NM 87545, United States

[3]Theoretical Division, Los Alamos National Laboratory, Los Alamos, NM 87545, United States

*Correspondence: kortkamp@lanl.gov

# S1. Quantification of voltage loss modes

To calculate the overall cell voltage, we consider the reversible potential and contributions stemming from the key voltage loss mechanisms, namely kinetic, ohmic, and mass transport losses, the details of which are elaborated below:

$$V = E^{rev} + |\eta_{kinetic}^{anode}| + |\eta_{kinetic}^{cathode}| + |\eta_{ohmic}| + |\eta_{mt}|. \tag{S1}$$

The kinetic losses take into account the activation overpotential, which occurs at the electrode-electrolyte interface and is evaluated by volume integration of the anodic and cathodic overpotentials at the respective electrodes.

$$\eta_{kinetic}^{anode} = \frac{1}{L_{ACL}} \int \eta_a dx \tag{S2}$$

$$\eta_{kinetic}^{cathode} = \frac{1}{L_{CCL}} \int \eta_c dx \tag{S3}$$

The ohmic loss mode ($\eta_{ohmic}$) considers the ionic resistance arising from the membrane and catalyst layers, in addition to the electronic resistance attributed to the catalyst layers, porous transport layers, and bipolar plates. It is computed as follows[1]:

$$\eta_{electronic} = \Delta\phi_e|_{PTL} + \Delta\phi_e|_{ACL} + \Delta\phi_e|_{CCL} + \Delta\phi_e|_{GDL}, \tag{S4}$$

$$\eta_{ionic} = \Delta\phi_p|_{ACL} + \Delta\phi_p|_{PEM} + \Delta\phi_p|_{CCL}, \tag{S5}$$

$$\eta_{ohmic} = \eta_{electronic} + \eta_{ionic}, \tag{S6}$$

where $\Delta\phi_{e/p}|_i$ represents the potential difference across the boundaries of the respective layers. The mass transport loss ($\eta_{mt}$) comprises the diffusion overpotential and the bubble overpotential. The diffusion overpotential ($\eta_{diffusion}$) originates from the slow transport of evolved gases through the electrodes. The bubble overpotential ($\eta_{bubble}$) dominates on the anode side owing to the accumulation of electrogenerated oxygen bubbles, which hinders the liquid water access. The expressions for both loss modes are given by[2]:

$$\eta_{diffusion} = \frac{RT}{4F} \ln\left(\frac{c_{O_2}}{c_{O_2,ref}}\right) + \frac{RT}{2F} \ln\left(\frac{c_{H_2}}{c_{H_2,ref}}\right), \tag{S7}$$

$$\eta_{bubble} = \frac{RT}{4F} \ln\left(\frac{1}{\left(1 - s_{O_2,ACL}\right)^{n_s}}\right), \tag{S8}$$

$$\eta_{mt} = \eta_{diffusion} + \eta_{bubble}, \tag{S9}$$

where $c_{O_2,ref}$ and $c_{H_2,ref}$ represent the reference oxygen and hydrogen concentrations, respectively. Further, $s_{O_2,ACL}$ indicates the average oxygen gas saturation at the anode catalyst layer. The OER coverage factor is denoted by the exponent, $n_s$.

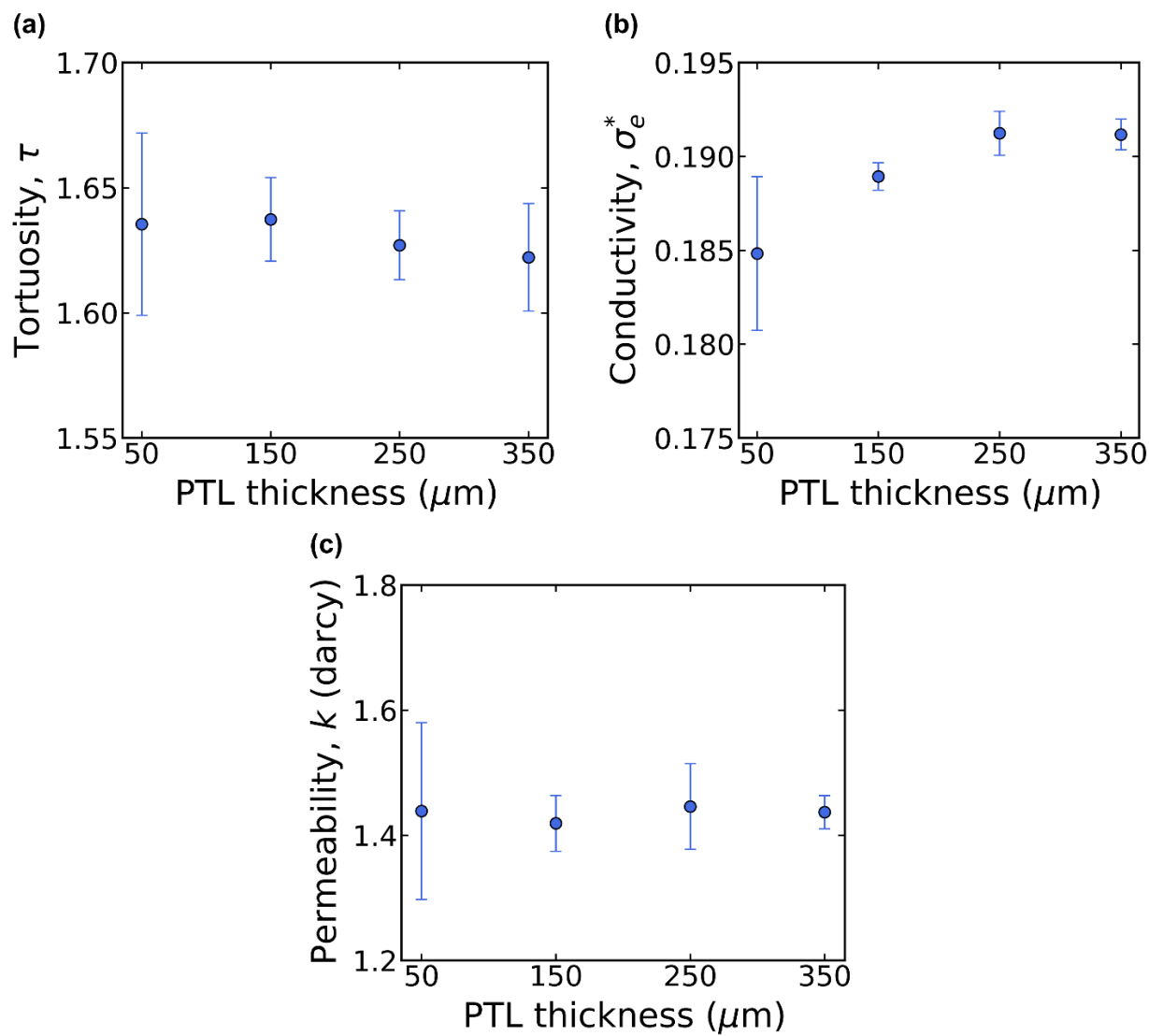

**Figure S1**: Influence of PTL thickness on (a) tortuosity ($\tau$), (b) normalized effective electronic conductivity ($\sigma_e^*$), and (c) single-phase (absolute) permeability ($k$) at an intermediate PTL porosity of 0.55. At each PTL thickness, PNM predictions from 10 stochastic microstructural realizations are shown. Symbols indicate mean values, while error bars represent the corresponding standard deviations.

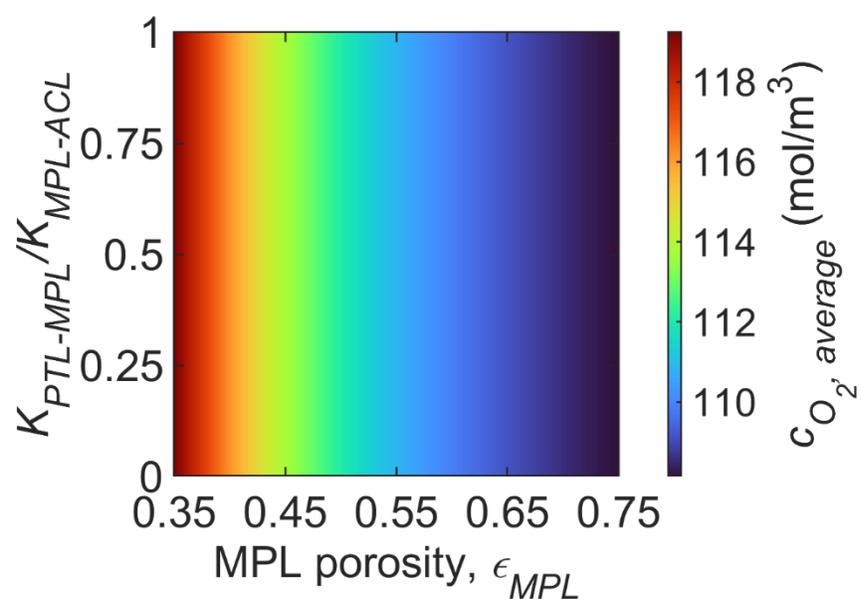

**Figure S2**: Effect of bilayer PTLs: interplay of the specific interfacial resistance ratio and MPL porosity on the average anode oxygen concentration at $I = 4$ A/cm$^2$.

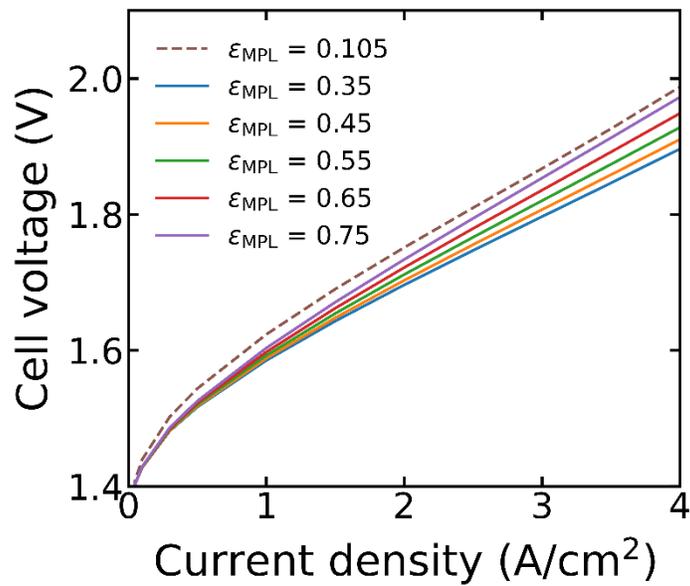

**Figure S3**: Effect of tuning the MPL porosity on electrochemical performance. The same porosity window employed for the single-layer PTLs is considered, in addition to an ultra-low porosity case ($\varepsilon_{MPL} = 0.105$) shown by dashed lines. The MPL thickness is fixed at 15 μm, and the backing layer porosity is 0.75.

Table S1. Fluxes corresponding to the variables solved in the multiphysics model

| Symbol | Expression | Description |
|---|---|---|
| $j_e$ | $-\sigma_e^{eff}\nabla\phi_e$ | Electronic flux |
| $j_p$ | $-\sigma_p^{eff}\nabla\phi_p$ | Protonic flux |
| $j_{O_2}$ | $-D_{O_2}^{eff}\nabla c_{O_2}$ | Oxygen flux |
| $j_{H_2O}$ | $-D_{H_2O}^{eff}\nabla c_{H_2O}$ | Water vapor flux |
| $j_{H_2}$ | $-D_{H_2}^{eff}\nabla c_{H_2}$ | Hydrogen flux |
| $j_{DW}$ | $-D_{DW}\varepsilon_{ionomer}^{1.5}\nabla c_{DW} + \dfrac{n_{drag}V_m c_{SO_3^-}}{F}j_p$ | Ionomer-dissolved water flux |
| $j_l$ | $-\dfrac{\rho_l k k_{rl}}{\mu_l}\nabla p_l$ | Liquid-phase mass flux |
| $j_g$ | $-\dfrac{\rho_g k k_{rg}}{\mu_g}\nabla p_g$ | Gas-phase mass flux |

**Table S2.** Anode inlet parameters used in the multiphysics model

| Symbol | Expression | Description |
|---|---|---|
| $x_{H_2O,a}$ | $\dfrac{RH_a c_{H_2O}^{sat}}{p_{g,a}^{in}}$ | Mole fraction of water vapor in anode gas channel |
| $x_{O_2,a}$ | $1 - x_{H_2O,a}$ | Mole fraction of oxygen in anode gas channel |
| $c_{O_2,a}$ | $\dfrac{x_{O_2,a} p_{g,a}^{in}}{RT}$ | Oxygen concentration at anode inlet |

**Table S3.** Cathode inlet parameters used in the multiphysics model

| Symbol | Expression | Description |
|---|---|---|
| $x_{H_2O,c}$ | $\dfrac{RH_c c_{H_2O}^{sat}}{p_{g,c}^{in}}$ | Mole fraction of water vapor in cathode gas channel |
| $x_{H_2,c}$ | $1 - x_{H_2O,c}$ | Mole fraction of hydrogen in cathode gas channel |
| $c_{H_2,c}$ | $\dfrac{x_{H_2,c} p_{g,c}^{in}}{RT}$ | Hydrogen concentration at cathode inlet |

**Table S4.** Boundary conditions used in the multiphysics model

| Variable | AGC/PTL | PTL/ACL | ACL/PEM | PEM/CCL | CCL/GDL | GDL/CGC |
|---|---|---|---|---|---|---|
| $\phi_e$ | $-\sigma_e^{eff}\nabla\phi_e = I$ | continuity | $n \cdot j_e = 0$ | $n \cdot j_e = 0$ | continuity | $\phi_e = 0$ |
| $\phi_p$ | — | $n \cdot j_p = 0$ | continuity | continuity | $n \cdot j_p = 0$ | — |
| $c_{O_2}$ | $c_{O_2,a}$ | continuity | $n \cdot j_{O_2} = 0$ | — | — | — |
| $c_{H_2O}$ | $RH_a c_{H_2O}^{sat}$ | continuity | $n \cdot j_{H_2O} = 0$ | $n \cdot j_{H_2O} = 0$ | continuity | $RH_c c_{H_2O}^{sat}$ |
| $c_{H_2}$ | — | — | — | $n \cdot j_{H_2} = 0$ | continuity | $c_{H_2,c}$ |
| $c_{DW}$ | — | $n \cdot j_{DW} = 0$ | continuity | continuity | $n \cdot j_{DW} = 0$ | — |
| $p_l$ | $p_{l,a}^{in}$ | continuity | $n \cdot j_l = 0$ | $n \cdot j_l = 0$ | continuity | $p_{l,c}^{in}$ |
| $p_g$ | $p_{g,a}^{in}$ | continuity | $n \cdot j_g = 0$ | $n \cdot j_g = 0$ | continuity | $p_{g,c}^{in}$ |

**Table S5.** Operating conditions used in the multiphysics model

| Symbol | Expression | Description |
|---|---|---|
| $p^{ref}$ | 1 atm | Reference pressure |
| $T^{ref}$ | 353.15 K | Reference temperature |
| $p_{op,a}$ | 1 atm | Operating pressure at the anode |
| $p_{l,a}^{in}$ | $p_{op,a} \times 1.01325 \times 10^5$ Pa | Liquid pressure in anode gas channel |
| $p_{g,a}^{in}$ | $p_{l,a}^{in} + 6100$ Pa | Gas pressure in anode gas channel |
| $p_{op,c}$ | 1 atm | Operating pressure at the cathode |
| $p_{g,c}^{in}$ | $p_{op,c} \times 1.01325 \times 10^5$ Pa | Gas pressure in cathode gas channel |
| $p_{l,c}^{in}$ | $p_{g,c}^{in}$ | Liquid pressure in cathode gas channel |
| $RH_a$ | 1 | Relative humidity in anode gas channel |
| $RH_c$ | 1 | Relative humidity in cathode gas channel |
| $T_{op}$ | 353.15 K | Operating temperature of the PEMWE |
| $I$ | *variable* | Operating current density |

**Table S6.** Thermodynamic parameters used in the multiphysics model

| Symbol | Expression | Description |
|---|---|---|
| $\Delta H_a$ | 285.83 J/mol | Enthalpy change of OER |
| $\Delta H_c$ | 0 J/mol | Enthalpy change of HER |
| $\Delta S_a$ | 163.3 Jmol$^{-1}$K$^{-1}$ | Entropy change of OER |
| $\Delta S_c$ | $-0.104$ Jmol$^{-1}$K$^{-1}$ | Entropy change of HER |
| $E_{ad}$ | $30 \times 10^3$ J/mol | Activation energy for adsorption/ desorption |
| $k_d^{ref}$ | 28 cm/s | Reference desorption mass transfer coefficient |
| $k_a^{ref}$ | $k_d^{ref}/10$ | Reference adsorption mass transfer coefficient |
| $a_{lg}$ | $3 \times 10^6$ m$^{-1}$ | Specific liquid-gas interfacial area |
| $k_e^{ref}$ | $5 \times 10^{-4}$ | Reference evaporation mass transfer coefficient |
| $k_c^{ref}$ | $6 \times 10^{-3}$ | Reference condensation mass transfer coefficient |

**Table S7.** Kinetic parameters used in the multiphysics model

| Symbol | Expression | Description |
|---|---|---|
| $a_a$ | $6\varepsilon_{IrO_2}/d_{anode}$ | Electrochemical active area of anode catalyst layer (ACL) |
| $a_c$ | $6(\varepsilon_{Pt} + \varepsilon_C)/d_{cathode}$ | Electrochemical active area of cathode catalyst layer (CCL) |
| $\alpha_a$ | 0.5 | Charge transfer coefficient of OER |
| $\alpha_c$ | 0.5 | Charge transfer coefficient of HER |
| $E_a$ | $4 \times 10^4$ J/mol | Activation energy for OER exchange current density |
| $E_c$ | $2 \times 10^4$ J/mol | Activation energy for HER exchange current density |
| $i_{0,a}^{ref}$ | $8 \times 10^{-3}$ A/m² (for MEAs with single-layer PTLs) $2.5 \times 10^{-2}$ A/m² (for MEAs with bilayer and trilayer PTLs) | Reference anode exchange current density |
| $i_{0,c}^{ref}$ | 50 A/m² | Reference cathode exchange current density |
| $n_s$ | 2 | OER coverage exponent |

**Table S8.** Anode parameters used in the multiphysics model

| Symbol | Expression | Description |
|---|---|---|
| $L_{PTL}$ | 250 µm | PTL thickness |
| $\varepsilon_{PTL}$ | variable (refer to Table 2 in the main text) | PTL porosity |
| $L_{ACL}$ | 0.5 µm | ACL thickness |
| $\rho_{IrO_2}$ | 11.66 g/cm³ | Density of $IrO_2$ catalyst |
| $m_{IrO_2}$ | 0.1 mg/cm² | Anode catalyst loading |
| $\omega_{ion}$ | 8.5% | Weight % of ionomer in the ACL |
| $m_{ion}$ | $m_{IrO_2}\omega_{ion}/(1-\omega_{ion})$ | Ionomer loading in the ACL |
| $\varepsilon_{IrO_2}$ | $m_{IrO_2}/\rho_{IrO_2}L_{ACL}$ | Volume fraction of $IrO_2$ catalyst in the ACL |
| $\varepsilon_{ionomer,ACL}$ | $m_{ion}/\rho_{wet-mem}L_{ACL}$ | Volume fraction of ionomer in the ACL |
| $\varepsilon_{ACL}$ | $1 - \varepsilon_{IrO_2} - \varepsilon_{ionomer,ACL}$ | Porosity of the ACL |
| $d_{anode}$ | 40 nm | Diameter of anode catalyst particles |
| $K_{PTL-ACL}$ | 80 mΩ cm² | Specific interfacial resistance at the PTL-ACL boundary |
| $\varepsilon_{MPL}$ | variable | MPL porosity |
| $L_{MPL}$ | 15 µm (unless otherwise mentioned) | MPL thickness |
| $K_{MPL-ACL}$ | 50 mΩ cm² | Specific interfacial resistance at the MPL-ACL boundary |

Table S9. Membrane parameters used in the multiphysics model

| Symbol | Expression | Description |
|---|---|---|
| $L_{PEM}$ | 125 μm | PEM thickness |
| $\varepsilon_{ionomer,PEM}$ | 1 | Volume fraction of ionomer in the PEM |
| $EW_{ionomer}$ | 1.1 kg/mol | Equivalent weight of ionomer |
| $\rho_{dry-mem}$ | $1.98 \times 10^3$ kg/m³ | Density of dry membrane |
| $V_m$ | $EW_{ionomer}/\rho_{dry-mem}$ | Molar volume of ionomer |
| $c_{SO_3^-}$ | $1/V_m$ | $SO_3^-$ concentration in the ionomer |
| $\lambda_{global}$ | $min(22, 0.043 + 17.81 RH_{local} - 39.85 RH_{local}^2 + 36 RH_{local}^3)$ | Water uptake (global) |
| $\rho_{wet-mem}$ | $\rho_{dry-mem} \dfrac{\left\{1 + \left(\dfrac{M_{H_2O}\lambda_{global}}{EW_{ionomer}}\right)\right\}}{\left\{1 + \left(\dfrac{M_{H_2O}\lambda_{global}\rho_{dry-mem}}{EW_{ionomer}\rho_{H_2O}}\right)\right\}}$ | Density of wet membrane |

**Table S10.** Cathode parameters used in the multiphysics model

| Symbol | Expression | Description |
|---|---|---|
| $L_{GDL}$ | 370 μm | GDL thickness |
| $\varepsilon_{GDL}$ | 0.78 | GDL porosity |
| $L_{CCL}$ | 9 μm | CCL thickness |
| $\rho_{Pt}$ | 21.45 g/cm$^3$ | Density of platinum |
| $\rho_C$ | 2 g/cm$^3$ | Density of carbon |
| $m_{Pt}$ | 0.1 mg/cm$^2$ | Cathode catalyst loading |
| $R_{C/Pt}$ | 1.82 | Carbon/platinum ratio |
| $R_{I/C}$ | 1.5 | Ionomer/carbon ratio |
| $\varepsilon_{Pt}$ | $m_{Pt}/\rho_{Pt}L_{CCL}$ | Volume fraction of platinum in the CCL |
| $\varepsilon_C$ | $m_{Pt}R_{C/Pt}/\rho_C L_{CCL}$ | Volume fraction of carbon in the CCL |
| $\varepsilon_{ionomer,CCL}$ | $m_{Pt}R_{C/Pt}R_{I/C}/\rho_{wet-mem}L_{CCL}$ | Volume fraction of ionomer in the CCL |
| $\varepsilon_{CCL}$ | $1-\varepsilon_{Pt}-\varepsilon_C-\varepsilon_{ionomer,CCL}$ | Porosity of the CCL |
| $d_{cathode}$ | 40 nm | Diameter of platinum/carbon particles |

**Table S11.** Tortuosity values used in the multiphysics model

| Symbol | Expression | Description |
|---|---|---|
| $\tau_{PTL}$ | variable (refer to Table 2 in the main text) | Tortuosity of PTL |
| $\tau_{ACL}$ | $\varepsilon_{ACL}^{-0.5}$ | Tortuosity of ACL |
| $\tau_{CCL}$ | $\varepsilon_{CCL}^{-0.5}$ | Tortuosity of CCL |
| $\tau_{GDL}$ | 1.242 | Tortuosity of GDL |
| $n_b$ | $= n_s$ | Pore blockage exponent |
| $d_{pore}$ | 50 nm | Mean pore diameter of the catalyst layers |

**Table S12.** Conductivity values used in the multiphysics model

| Symbol | Expression | Description |
|---|---|---|
| $\sigma_{e,PTL}^{bulk}$ | $2.38 \times 10^6$ S/m | Bulk electronic conductivity of PTL |
| $\sigma_{e,ACL}^{bulk}$ | $2 \times 10^3$ S/m | Bulk electronic conductivity of ACL |
| $\sigma_{e,CCL}^{bulk}$ | $7.14 \times 10^4$ S/m | Bulk electronic conductivity of CCL |
| $\sigma_{e,GDL}^{bulk}$ | $1.25 \times 10^3$ S/m | Bulk electronic conductivity of GDL |
| $\sigma_{e,PTL}^{*}$ | variable (refer to Table 2 in the main text) | Normalized effective electronic conductivity of PTL |
| $\sigma_{e,ACL}^{*}$ | $\varepsilon_{IrO_2}^{1.5}$ | Normalized effective electronic conductivity of ACL |
| $\sigma_{e,CCL}^{*}$ | $(\varepsilon_{Pt} + \varepsilon_C)^{1.5}$ | Normalized effective electronic conductivity of CCL |
| $\sigma_{e,GDL}^{*}$ | 0.068 | Normalized effective electronic conductivity of GDL |
| $\sigma_{e,PTL}$ | $\sigma_{e,PTL}^{bulk} \times \sigma_{e,PTL}^{*}$ | Effective electronic conductivity of PTL |
| $\sigma_{e,ACL}$ | $\sigma_{e,ACL}^{bulk} \times \sigma_{e,ACL}^{*}$ | Effective electronic conductivity of ACL |
| $\sigma_{e,CCL}$ | $\sigma_{e,CCL}^{bulk} \times \sigma_{e,CCL}^{*}$ | Effective electronic conductivity of CCL |
| $\sigma_{e,GDL}$ | $\sigma_{e,GDL}^{bulk} \times \sigma_{e,GDL}^{*}$ | Effective electronic conductivity of GDL |
| $\sigma_{e,MPL}^{bulk}$ | $\sigma_{e,PTL}^{bulk}$ | Bulk electronic conductivity of MPL |
| $\sigma_{e,MPL}^{*}$ | variable | Normalized effective electronic conductivity of MPL |
| $\sigma_{e,MPL}$ | $\sigma_{e,MPL}^{bulk} \times \sigma_{e,MPL}^{*}$ | Effective electronic conductivity of MPL |

**Table S13.** Single-phase permeability values used in the multiphysics model

| Symbol | Expression | Description |
|---|---|---|
| $k_{PTL}$ | *variable* (refer to Table 2 in the main text) | Single-phase permeability of PTL |
| $k_{ACL}$ | $4 \times 10^{-14}$ m² | Single-phase permeability of ACL |
| $k_{CCL}$ | $4 \times 10^{-14}$ m² | Single-phase permeability of CCL |
| $k_{GDL}$ | $5.295 \times 10^{-12}$ m² | Single-phase permeability of GDL |

**Table S14.** Two-phase transport parameters used in the multiphysics model

| Symbol | Expression | Description |
|---|---|---|
| $s_r$ | 0.08 | Residual saturation |
| $s_m$ | 1 | Maximum saturation |
| $s_{im}$ | 0.1 | Immobile liquid saturation |

**Table S15.** Van-Genuchten fitting parameters used in the multiphysics model

| Parameter | PTL | ACL | CCL | GDL |
|---|---|---|---|---|
| $f_1$ | 1 | 1 | 0.4 | 0.4 |
| $f_2$ | 0 | 0 | 0.6 | 0.6 |
| $m_1$ | 400 | 400 | 125 | 125 |
| $n_1$ | 0.3 | 0.3 | 0.9 | 0.9 |
| $m_2$ | — | — | 150 | 150 |
| $n_2$ | — | — | 1.5 | 1.5 |
| $p_{cb,1}$ | 107500 Pa | 108000 Pa | 100500 Pa | 101500 Pa |
| $p_{cb,2}$ | — | — | 104950 Pa | 106000 Pa |

**Table S16.** Chemical and thermophysical properties used in the multiphysics model

| Symbol | Expression | Description |
|---|---|---|
| $M_{H_2O}$ | 18 g/mol | Molar mass of water |
| $M_{O_2}$ | 32 g/mol | Molar mass of oxygen |
| $M_{H_2}$ | 2 g/mol | Molar mass of hydrogen |
| $\mu_g$ | $2.03 \times 10^{-5}$ kgm⁻¹s⁻¹ | Dynamic viscosity of gas phase |
| $\mu_l$ | $10^{-3} exp\left(-3.63148 + \frac{542.05}{T-144.15}\right)$ kgm⁻¹s⁻¹ | Dynamic viscosity of liquid water |
| $\rho_{H_2O}$ | 1000 kg/m³ | Density of liquid water |

Table S17. Diffusion coefficients used in the multiphysics model

| Symbol | Expression | Description |
|---|---|---|
| $D_{H_2O,O_2}$ | $0.36 \times 10^{-4} \left(\frac{T}{T^{ref}}\right)^{1.5} \left(\frac{p^{ref}}{p_{g,a}}\right)$ m²/s | Binary diffusion coefficient of water at the anode |
| $D_{H_2O,H_2}$ | $1.24 \times 10^{-4} \left(\frac{T}{T^{ref}}\right)^{1.5} \left(\frac{p^{ref}}{p_{g,c}}\right)$ m²/s | Binary diffusion coefficient of water at the cathode |
| $D_{H_2O,k}$ | $\left(\frac{d_{pore}}{3}\right)\sqrt{\frac{8RT}{\pi M_{H_2O}}}$ | Knudsen diffusion coefficient of water |
| $D_{O_2,k}$ | $\left(\frac{d_{pore}}{3}\right)\sqrt{\frac{8RT}{\pi M_{O_2}}}$ | Knudsen diffusion coefficient of oxygen |
| $D_{H_2,k}$ | $\left(\frac{d_{pore}}{3}\right)\sqrt{\frac{8RT}{\pi M_{H_2}}}$ | Knudsen diffusion coefficient of hydrogen |
| $D_{H_2O,a}$ | $D_{H_2O,O_2}$ (APTL) <br> $\left(\frac{1}{D_{H_2O,O_2}} + \frac{1}{D_{H_2O,k}}\right)^{-1}$ (ACL) | Combined diffusion coefficient of water at the anode |
| $D_{H_2O,c}$ | $D_{H_2O,H_2}$ (GDL) <br> $\left(\frac{1}{D_{H_2O,H_2}} + \frac{1}{D_{H_2O,k}}\right)^{-1}$ (CCL) | Combined diffusion coefficient of water at the cathode |
| $D_{O_2}$ | $D_{H_2O,O_2}$ (APTL) <br> $\left(\frac{1}{D_{H_2O,O_2}} + \frac{1}{D_{O_2,k}}\right)^{-1}$ (ACL) | Combined diffusion coefficient of oxygen |
| $D_{H_2}$ | $D_{H_2O,H_2}$ (GDL) <br> $\left(\frac{1}{D_{H_2O,H_2}} + \frac{1}{D_{H_2,k}}\right)^{-1}$ (CCL) | Combined diffusion coefficient of hydrogen |